\documentclass[prd,amsmath,floats,amssymb, floatfix,superscriptaddress,nofootinbib,onecolumn]{revtex4} 
\usepackage[plainpages=false, colorlinks=true, anchorcolor=blue, linkcolor=blue, citecolor=blue, bookmarks=false]{hyperref}

\usepackage[T1]{fontenc}
\usepackage{color}
\usepackage{graphicx}
\usepackage{dcolumn}
\usepackage{multirow}
\usepackage{bm}
\usepackage{longtable}
\usepackage{adjustbox}
\newcommand{\rthis}[1]{\textcolor{black}{#1}}
\usepackage{float}
\usepackage[caption=false]{subfig}
\usepackage[paperwidth=210mm,paperheight=297mm,centering,hmargin=2cm,vmargin=2.5cm]{geometry}
\usepackage{appendix}

\begin{document}

\title{A MINOT-based Study of Gamma-ray emission from SPT-CL J2012-5649/Abell 3667}

\author{Siddhant Manna}
 \altaffiliation{Email:ph22resch11006@iith.ac.in}
\author{Shantanu Desai}
 \altaffiliation{Email:shntn05@gmail.com}
\affiliation{
 Department of Physics, IIT Hyderabad Kandi, Telangana 502284, India}

\begin{abstract}
We present an analysis of the non-thermal properties of the merging galaxy cluster SPT-CL J2012-5649/Abell~3667 ($z = 0.0556$, $M_{500} = 7.16 \times 10^{14}\
M_\odot$) using the MINOT non-thermal emission modeling
framework. The predicted hadronic gamma-ray flux from $pp$ interactions
in the $1$--$300\ \mathrm{GeV}$ band is $2.82 \times
10^{-11}\ \mathrm{cm^{-2}\ s^{-1}}$ within $R_{500}$,
rising to $1.15 \times 10^{-10}\ \mathrm{cm^{-2}\ s^{-1}}$
at the truncation radius ($3.7\,R_{500}$), in
broad agreement with the Fermi-LAT
reported flux of $1.3 \times 10^{-10}\ \mathrm{cm^{-2}\
s^{-1}}$. Approximately $76\%$ of the predicted hadronic
flux originates from beyond $R_{500}$. The Inverse Compton   contribution from cosmic-ray electrons is subdominant relative to the hadronic $\pi^0$-decay gamma-ray component by a factor of ${\sim}20$ in the $1$--$300\ \mathrm{GeV}$ energy band, and therefore does not contribute significantly to the observable signal.
\rthis{The best fit of the observed Fermi-LAT data is achieved with a hadronic emission assuming a proton spectrum $\propto E^{-3.5}$ and a proton energy density almost in equipartition with the thermal energy density. Even though those parameters may look extreme, they could well be explained by the particle acceleration by low Mach number accretion shocks.}
\end{abstract}


\maketitle

\section{\label{sec:intro}Introduction}
Galaxy clusters are formed through the gravitational collapse of overdense regions that existed in the early universe. Over cosmic time, density perturbations cause these regions to gravitationally attract surrounding matter, gradually building up into clumps and filaments that eventually coalesce into clusters \citep{Rees1978}. As a result, galaxy clusters represent the largest structures in the universe that are both gravitationally bound and virialized, making them exceptional laboratories for investigating cosmology \citep{Kravtsov2012, Allen2011, Vikhlininrev2014,Miyatake25} and fundamental physics~\citep{Bohringer2016,Desai2018, Bora2021, Bora2021b, Bora2022}. Observationally, galaxy clusters have been detected across a broad range of wavelengths, spanning from radio \citep{Feretti2012} to hard X-rays \citep{Wik2014}. Over the last two decades, dedicated surveys conducted in the optical, X-ray, and microwave bands have significantly expanded the known cluster population, enabling a wide variety of cosmological and astrophysical investigations, often through multi-wavelength analyses.

Although initial searches for gamma-ray emission from galaxy clusters  using the Fermi-LAT telescope reported null results~\cite{Ackermann2010,Ackermann2014,Ackermann2016}, in recent years several groups have found a gamma-ray signal from clusters.
In \citep{Manna2024},  we performed a systematic search for gamma-ray emission from 300 galaxy clusters selected from the SPT-SZ survey using approximately 15 years of Fermi-LAT data in the $1$--$300$~GeV energy range. Our analysis reported a $\sim6.1\sigma$ gamma-ray detection from the cluster SPT-CL~J2012$-$5649, coincident with Abell 3667 with emission primarily observed between $1$--$10$~GeV. However, because several radio galaxies lie within the Fermi-LAT point spread function around the cluster, the study could not conclusively establish the cluster origin of the observed gamma-ray signal. Subsequently, \citet{Harale2025} reported a $\sim4\sigma$ excess of diffuse gamma-ray emission from the dynamically active galaxy cluster Abell~119 using 14 years of Fermi-LAT data. The detected excess was spatially offset from the cluster center and was found to be better described by an extended emission model, suggesting a possible origin from nonthermal hadronic processes in the intracluster medium. Recently, \citet{Li2026} carried out a systematic analysis of 65 galaxy clusters using 16 years of Fermi-LAT observations in the $0.5$--$500$~GeV range and reported significant gamma-ray excesses from the clusters Abell 2065 and Abell 2244 with local significances of $\sim5\sigma$. Their study concluded that the detected emission is unlikely to originate from dark matter annihilation and is more plausibly associated with hadronic cosmic-ray interactions within the intracluster medium. They also reported a possible gamma-ray excess from the merging galaxy cluster Abell 3667 with TS $\sim 24.3$ detected near the cluster center, with an integral photon flux of $(2.93 \pm 0.69)\times10^{-10}\ \mathrm{ph\ cm^{-2}\ s^{-1}}$ in the $0.5$--$500\ \mathrm{GeV}$ energy range. However, they also concluded that the  observed emission may be associated with nearby radio sources, and therefore the cluster origin of the gamma-ray signal remains uncertain.

In a series of studies based on stacked Fermi-LAT observations of galaxy clusters, U. Keshet ~\cite{ReissKeshet2018,Keshet2025a,Keshet2025b} reported statistically significant gamma-ray excesses associated with both cluster virial shocks and central regions. The 2018 study \citep{ReissKeshet2018} detected a gamma-ray ring at the virial shock radius of stacked clusters with a significance of $\sim5.8\sigma$, interpreted as Inverse Compton (IC)  emission from relativistic electrons accelerated at virial shocks. More recent analyses \citep{Keshet2025a,Keshet2025b} reported extended central gamma-ray excesses consistent with hadronic cosmic-ray interactions in the intracluster medium, along with excess gamma-ray sources near cluster virial radii potentially associated with radio relics and merger-driven substructures. Subsequently, we also carried out a stacked analysis of SPT-SZ galaxy clusters using 16.4 years of Fermi-LAT observations and reported a statistically significant cumulative gamma-ray signal with TS $=75.2$, corresponding to a significance of $\sim8.4\sigma$~\cite{Mannastacked}. The derived stacked spectrum is well described by a power-law with spectral index $-2.59\pm0.20$. Our study suggested that the strong cumulative signal is likely dominated by AGN-containing clusters, whereas the lower-TS cluster subsample shows characteristics broadly consistent with diffuse hadronic emission from the intracluster medium.

Several physical mechanisms capable of producing gamma-ray emission within galaxy clusters have been proposed in the literature, which we briefly summarize below. Galaxy clusters harbor dense concentrations of galaxies, non-baryonic dark matter (comprising roughly 80\% of the total mass), and hot diffuse gas (accounting for approximately 10--15\%). They also serve as vast reservoirs of high-energy relativistic cosmic rays (CRs),  both electrons and protons permeating the hot, ionized intracluster medium (ICM)~\citep{Brunetti2014, Wittor2023}. Observational evidence for the acceleration of cosmic ray electrons is provided by radio relics detected within merging clusters~\citep{Paul2023,Wittor2023}, which are believed to arise from merger-driven shock waves capable of boosting particles to extreme energies. Such energetic particles can give rise to gamma-ray emission through several channels, including IC scattering of relativistic electrons off CMB photons, non-thermal bremsstrahlung radiation, and the decay of neutral pions produced when cosmic ray protons interact with ICM gas \citep{Ensslin1997,Hinton2007,Vazza2016,Petrosian2001,Brunetti2017}. In a recent study, the cumulative gamma-ray flux from galaxy clusters was estimated by coupling cosmological MHD simulations with cosmic ray transport modeling up to redshift $z \leq 5$, revealing that this integrated flux could account for as much as 100\% of the diffuse gamma-ray background at those redshifts \citep{Saqib2023}. Given that the dominant mass component in galaxy clusters is cold dark matter, gamma-ray signals could also arise from the annihilation of dark matter WIMPs within cluster halos \citep{Ackermann2010,DiMauro2023,Murase2023}. In addition to ICM-related emission processes, gamma rays may also originate from star-forming activity occurring in galaxies that are members of the cluster \citep{Storm2012}.

Both classes of emission require populations of ultra-relativistic
cosmic ray electrons (CRe) emitting
synchrotron radiation in the $\mu\mathrm{G}$-level magnetic
fields of the ICM. The spectral and morphological properties
of the radio emission, combined with X-ray observations of
the thermal gas, provide constraints on the magnetic field
strength and the CRe energy distribution
but cannot alone distinguish between leptonic (direct CRe
acceleration) and hadronic (CRe produced as secondary products
of cosmic ray protons (CRp)--ICM proton interactions) origins for the emitting
particles \citep{Blasi1999, Pfrommer2008, Brunetti2017}.

Gamma-ray observations offer a powerful complementary probe
of the non-thermal ICM. Hadronic interactions between
CRp and thermal protons produce neutral pions that decay into
gamma-ray photon pairs ($p + p \to \pi^0 \to \gamma + \gamma$),
generating a high-energy signal that scales as the product of
the CRp and gas densities integrated over the cluster volume
\citep{Blasi1999, Miniati2003, Pinzke2010}. The predicted
hadronic gamma-ray flux therefore provides a direct handle
on the CRp energy content and spatial distribution in the ICM.
In addition, IC scattering of CRe off the
cosmic microwave background (CMB) produces gamma-ray emission
at energies accessible to the Fermi Large Area Telescope
(Fermi-LAT) providing an independent
constraint on the leptonic CR population
\citep{Rephaeli1979, Reimer2004, Brunetti2017}.

The non-detections from gamma-ray searches using Fermi-LAT in the first decade after launch~\cite{Ackermann2010,Ackermann2014,Ackermann2016} place upper limits on the CRp-to-thermal energy ratio of order $X_{\rm CRp}
\lesssim 10^{-2}$, broadly consistent with predictions from
cosmological simulations of structure formation~\citep{Pinzke2010, Vazza2016}. These upper limits motivate
detailed, cluster-by-cluster modeling of the expected
non-thermal emission, particularly for systems with
well-characterized X-ray thermodynamic structure and
prominent non-thermal radio features. 

Abell~3667 (hereafter A3667) is one of the most compelling
targets for such a study, given that gamma-ray emission with $\geq 5\sigma$ significance was recently detected~\cite{Manna2024}. It is a massive ($M_{500} =
7.16 \times 10^{14}\ M_\odot$), dynamically disturbed
galaxy cluster at redshift $z = 0.0556$, undergoing a
major binary merger~\citep{Owers2009}. A3667 hosts prominent merger-driven shocks and double radio
relic systems \citep{JohnstonHollitt2008, Hindson2014,Omiya2024,deGasperin2022},
with two large-scale arc-shaped relics. The cluster
also exhibits prominent X-ray cold fronts~\citep{Vikhlinin2001},
providing direct evidence for complex merger history. The combination of a well-studied radio relic system, deep archival X-ray data, and the ongoing merger morphology makes A3667 an ideal laboratory for
multiwavelength non-thermal emission modeling.

Multi-band observational constraints on A3667 now span
from hard X-rays to GeV gamma-rays. Fermi-LAT
observations have reported a gamma-ray signal in the
$1$--$300\ \mathrm{GeV}$ band with photon flux
${\sim}1.3 \times 10^{-10}\ \mathrm{cm^{-2}\ s^{-1}}$
and a steep spectral index $\Gamma = -3.61 \pm 0.33$
\citep{Manna2024,ShangLi}. However, the LAT emission is confined within $\sim 0.2\,R_{200}$ and detected only in the $1$--$10\ \mathrm{GeV}$ band~\cite{Manna2024}. Yet, MeerKAT data reveal several compact radio sources within a few arcminutes of the cluster center~\cite{deGasperin2022}. To clarify the nature of the emission,
multi-wavelength and multi-instrument follow-up studies have
been pursued. Archival COMPTEL data show no evidence of
MeV emission~\cite{Manna2024b}, while INTEGRAL constrains
the hard X-ray and soft gamma-ray emission in the
$3$--$300\ \mathrm{keV}$ band \citep{Manna2025i}. A
dedicated analysis with the DArk Matter Particle Explorer
(DAMPE)~\cite{Chang2017,Chang2014,Ambrosi2019} using
point-source, radial-disk, and radial-Gaussian models over
$3\ \mathrm{GeV}$--$1\ \mathrm{TeV}$ likewise revealed no
significant signal within $R_{200}$, yielding $95\%$ C.L.\
upper limits of
$\sim10^{-6}$--$10^{-4}\ \mathrm{MeV\ cm^{-2}\ s^{-1}}$
consistent with LAT measurements~\cite{Manna2024c}.

For A3667, previous hard X-ray studies have reported fluxes in the range of $\sim10^{-12}$--$10^{-11}$~erg~cm$^{-2}$~s$^{-1}$. Using combined \textit{XMM-Newton} and \textit{Swift}/BAT observations, \citet{Ajello2010} showed that the 50--100~keV emission can be modeled either by a hot thermal component or by a power law with photon index $\Gamma \approx 1.8$, corresponding to a flux of $3.0^{+4.2}_{-0.7}\times10^{-12}$~erg~cm$^{-2}$~s$^{-1}$. More recent deep \textit{NuSTAR} observations of the cluster core~\citep{Mirakhor2025} reported a 20--80~keV flux of $9.5^{+0.5}_{-5.5}\times10^{-12}$~erg~cm$^{-2}$~s$^{-1}$ for a power-law component with $\Gamma = 1.6^{+0.1}_{-0.2}$, although the data were found to favor a multi-temperature thermal model. The brightest point source resolved by \textit{Chandra} contributes only $2.2^{+1.5}_{-1.1}\times10^{-13}$~erg~cm$^{-2}$~s$^{-1}$ in the same band, suggesting that the observed hard X-ray excess is unlikely to be dominated by a single unresolved AGN. Similarly, \textit{Suzaku} observations in the 0.5--40~keV band found that the emission above 10~keV near the cluster center is better explained by a very hot thermal component ($kT>13$~keV) rather than IC emission. No statistically significant non-thermal emission was detected from the northwestern radio relic, yielding a 90\% confidence upper limit of $7.3\times10^{-13}$~erg~cm$^{-2}$~s$^{-1}$ in the 10--40~keV band \citep{Nakazawa2009}.

In this work, we present a model for A3667 using the
MINOT framework \citep[modeling and Interpretation of Non-Thermal emission from galaxy clusters Of the milky way Type;][]{Adam2020},
a publicly available code designed for self-consistent
modeling of non-thermal cluster emission across wavelengths. 
The MINOT framework has also been used to model the expected non-thermal hadronic gamma-ray and neutrino emission from galaxy clusters. For example, \citet{Voitsekhovskyi2021} studied the Hercules cluster (A2151), focusing on the brightest subclumps. Using thermal X-ray observations as inputs to the MINOT code, they simulated the hadronic gamma-ray and neutrino emission produced through cosmic-ray proton interactions in the intracluster medium. Assuming typical cosmic-ray parameters, namely a cosmic-ray-to-thermal energy ratio of $X_{\rm CR,p}\approx0.04$--$0.06$ and a proton spectral index $\gamma \approx 2.5$, the predicted gamma-ray and neutrino fluxes were found to remain below the sensitivity limits of current and planned instruments. However, the authors showed that additional cosmic-ray injection from AGN activity, mergers, accretion flows, and turbulence could increase the cosmic-ray energy fraction up to $X_{\rm CR,p}\approx0.1$ and produce a harder proton spectrum ($\gamma \sim 1.5$--$2.0$). Under these conditions, the brighter subclump could become detectable by CTA at the $\sim5\sigma$ level, while the associated hard-spectrum neutrino emission may also be detectable with next-generation observatories such as IceCube-Gen2.
We use the thermal electron density and pressure profiles of A3667 derived from
archival \textit{Chandra} X-ray data in the ACCEPT
catalog~\citep{Cavagnolo2009}, model the magnetic field
and cosmic-ray spatial distributions following standard
prescriptions, and compute the predicted hadronic and
leptonic gamma-ray emission for a range of CR spectral
and spatial model assumptions. We compare these predictions
with the available multi-instrument observational constraints
to assess the consistency of the hadronic model and quantify
the relative contribution of hadronic and leptonic emission
channels.

This paper is structured as follows. Section~\ref{sec:data_analysis}
describes the cluster properties, the ACCEPT thermodynamic
data, the profile construction methodology, and the MINOT
initialization including the non-thermal physics setup.
Section~\ref{sec:results} presents the predicted
thermodynamic structure, gamma-ray spectra, surface
brightness profiles, enclosed fluxes, and IC emission. The derived thermodynamic profiles
are collected in Appendix~\ref{app:thermodynamic}. Throughout
this work we adopt a flat $\Lambda$CDM cosmology with
$H_0 = 67.74\ \mathrm{km\,s^{-1}\,Mpc^{-1}}$ and
$\Omega_m = 0.3075$~\citep{Planck2015}.

\section{Data Analysis}
\label{sec:data_analysis}

\subsection{Cluster Properties and Cosmological Framework}
\label{sec:cluster_props}
At the cluster redshift, the corresponding angular diameter distance is $D_{\rm ang}=230.1\ \mathrm{Mpc}$. The adopted cluster mass is $M_{500}=7.16\times10^{14}\ M_\odot$, derived from the SPT-SZ catalog~\citep{Bleem2015}. The same Planck~2015 cosmology is used consistently throughout all profile derivations and MINOT modeling steps, following the cosmological framework adopted internally by MINOT~\citep{Adam2020}.

The characteristic radius $R_{500}$, defined as the radius enclosing a mean density 500 times the critical density of the Universe at the cluster redshift
$\rho_c(z)$, is computed as:
\begin{equation}
    R_{500} = \left(\frac{3\,M_{500}}{4\pi \cdot 500 \cdot
    \rho_c(z)}\right)^{1/3},
    \label{eq:R500}
\end{equation}
yielding $R_{500} = 1365.5\ \mathrm{kpc}$. The cluster center is located at $(\alpha,\,\delta)_{\rm J2000} = (303.1763^{\circ},\,
-56.8432^{\circ})$, as determined from the SPT-SZ catalogue \citep{Bleem2015}. A truncation radius of $R_{\rm trunc} = 5000\ \mathrm{kpc} \approx 3.7\,R_{500}$ is adopted. A
summary of the adopted cluster parameters is provided in
Table~\ref{tab:cluster_params}.

\begin{table}[ht]
\centering
\caption{Adopted global parameters for A3667 used throughout
this work.}
\label{tab:cluster_params}
\begin{tabular}{lcc}
\hline\hline
Parameter & Value & Unit \\
\hline
Redshift $z$              & $0.0556$                        & ---  \\
$(\alpha,\delta)_{\rm J2000}$ & $(303.1763, -56.8432)$      & deg  \\
$M_{500}$                 & $7.16 \times 10^{14}$           & $M_\odot$ \\
$R_{500}$                 & $1365.5$                 & kpc  \\
$R_{\rm trunc}$           & $5000$ & kpc  \\
$D_{\rm ang}$             & $230.1$                         & Mpc  \\
$H_0$                     & $67.74$                         & $\mathrm{km\,s^{-1}\,Mpc^{-1}}$ \\
$\Omega_m$                & $0.3075$                        & ---  \\
$E(z)$                    & $1.0267$                        & ---  \\
\hline
\end{tabular}
\end{table}

\subsection{ACCEPT Thermodynamic Data}
\label{sec:accept_data}
Thermodynamic profiles of A3667 are obtained from the ACCEPT
catalog~\citep{Cavagnolo2009}, which provides deprojected,
azimuthally averaged radial profiles derived from archival
\textit{Chandra} observations. The dataset combines 
\textit{Chandra} observations from four observing cycles (ObsIDs 5751, 5752, 5753, and 889),
yielding a total cleaned exposure time of ${\sim}343\ \mathrm{ks}$. We note that although this cluster has also been imaged using the eROSITA satellite~\cite{Bulbul24}, the eROSITA thermodynamic profiles are not yet publicly available (J. Sanders, private communication) and hence have not been used for this analysis.

The resulting profiles consist of $N = 56$ radial bins spanning
$2.7$--$295.2\ \mathrm{kpc}$, corresponding to
$0.002$--$0.216\,R_{500}$. Each bin provides the following data:
\begin{itemize}

    \item \textbf{Electron number density} $n_e(r)$
    [$\mathrm{cm^{-3}}$], derived from the spectral normalization
    of a single-temperature \texttt{APEC} plasma model. Across the
    56 radial bins,  the density spans $1.41 \times 10^{-3}$ to
    $5.73 \times 10^{-3}\ \mathrm{cm^{-3}}$, decreasing outward within the probed radial range. 
    
    \item \textbf{Electron temperature} $T(r)$ [keV], derived from
    the spectral shape of the \texttt{APEC} model fit. The observed
    temperature spans $5.08$--$6.01\ \mathrm{keV}$ across the
    $2.7$--$295.2\ \mathrm{kpc}$ radial range, with a median value
    of $5.5\ \mathrm{keV}$. The modest radial variation
    ($\lesssim 16\%$ peak-to-peak) is consistent with the limited
    radial coverage of the ACCEPT data ($r \lesssim 0.22\,R_{500}$). 

    \item \textbf{Electron thermal pressure} $P(r) = n_e\,k_{\rm B}T$
    [$\mathrm{dyne\ cm^{-2}}$], converted to
    $\mathrm{keV\,cm^{-3}}$ via $1\ \mathrm{dyne\ cm^{-2}} =
    6.242 \times 10^{8}\ \mathrm{keV\,cm^{-3}}$. The pressure
    spans $7.54 \times 10^{-3}$ to $2.91 \times 10^{-2}\
    \mathrm{keV\,cm^{-3}}$.
    
\end{itemize}

The ACCEPT profiles probe only the cluster core and inner
intracluster medium (ICM), extending to $0.216\,R_{500}$. Beyond
this radius, thermodynamic profiles must be extrapolated using
physically motivated parametric models, as described in
Section~\ref{sec:profiles}. We note that the temperature data are
directly used  in the construction of the self-consistent pressure
profile (cf. Section~\ref{sec:pressure}).

\subsection{Profile Construction and Extrapolation}
\label{sec:profiles}
The observed thermodynamic profiles are characterized  using parametric models and extrapolated to $R_{\rm trunc}=5000\ \mathrm{kpc}$ on a logarithmically spaced radial grid of 150 points spanning
$1$--$5500\ \mathrm{kpc}$. 

\subsubsection{Electron Density Profile}
\label{sec:density}

The electron density profile is modeled using the standard
$\beta$-model \citep{Cavaliere1976}:
\begin{equation}
    n_e(r) = n_0 \left[1 +
    \left(\frac{r}{r_c}\right)^2\right]^{-3\beta/2},
    \label{eq:betamodel}
\end{equation}
with all three parameters --- central density $n_0$,
core radius $r_c$, and slope $\beta$ treated as free parameters during the
profile fitting procedure. The best-fit parameters, their $1\sigma$ uncertainties, and the goodness-of-fit are summarized in Table~\ref{tab:fitparams}. The
low $\beta = 0.329$ reflects a shallow outer density gradient. The resulting density profile is shown
in the combined figure~\ref{fig:combined_profiles}, where the model is
extrapolated from the observed range ($r \lesssim 295\
\mathrm{kpc}$) to the truncation radius using the best-fit
$\beta$-model parameters.

\subsubsection{Thermal Pressure Profile}
\label{sec:pressure}
We construct the electron thermal pressure profile directly from the fundamental thermodynamic relation,
\begin{equation}
P_e(r)=n_e(r)\,k_{\rm B}\,T(r),
\end{equation}
where \(n_e(r)\) is the electron number density obtained from the best-fit
$\beta$-model. Following the convention adopted in MINOT, the electron pressure profile is used as the thermal pressure input. The corresponding
gas (proton) density required for the hadronic gamma-ray calculations is inferred from the measured electron density assuming a fully ionized intracluster plasma and $T(r)$ is derived from the ACCEPT
data as $T = P_{\rm ACCEPT}/n_{e,\rm ACCEPT}$, then extrapolated
using a power-law model:
\begin{equation}
    T(r) = T_0 \left(\frac{r}{r_0}\right)^{-\alpha},
    \label{eq:temperature_fit}
\end{equation}
with best-fit parameters $T_0 = 5.55\ \mathrm{keV}$,
$r_0 = 158.3\ \mathrm{kpc}$, and $\alpha = 0.039$
(Table~\ref{tab:fitparams}).

\begin{table}[ht]
\centering
\caption{Best-fit parameters for the electron density
($\beta$-model), and temperature power-law profiles of A3667, derived from
\textit{Chandra} ACCEPT data. The pressure profile adopted for MINOT modeling is computed as $P = n_e \times T$ using the $\beta$-model density and the power-law temperature fit.}
\label{tab:fitparams}
\begin{tabular}{llcc}
\hline\hline
Profile & Parameter & Value & Unit \\
\hline
\multirow{4}{*}{$\beta$-model (density)}
  & $n_0$        & $(4.49 \pm 0.15)\times10^{-3}$   & $\mathrm{cm^{-3}}$ \\
  & $r_c$        & $95.5 \pm 9.6$                   & kpc \\
  & $\beta$      & $0.329 \pm 0.019$                & --- \\
\hline
\multirow{4}{*}{Power-law (temperature)}
  & $T_0$        & $5.55$                           & keV \\
  & $r_0$        & $158.3$                          & kpc \\
  & $\alpha$     & $0.039$                          & --- \\
  & $T$ range    & $5.08$--$6.01$ (data)            & keV \\
\hline \hline
\end{tabular}
\end{table}

\subsection{MINOT Cluster Initialisation and Non-Thermal
            Physics Setup}
\label{sec:minot_init}
The thermodynamic profiles derived in Section~\ref{sec:profiles} are ingested into the MINOT framework \citep{Adam2020} to construct a complete physical model of A3667 for non-thermal emission modeling. The cluster object is initialized with the parameters summarized in
Tables~\ref{tab:cluster_params} and \ref{tab:fitparams}. Three additional physical components must be specified: the magnetic field profile, the cosmic-ray (CR) spatial distribution, and the CR energy spectra. All non-thermal parameters are collected in Table~\ref{tab:nonthermal_params}.

\subsubsection{Magnetic Field Profile}
\label{sec:magfield}
The ICM magnetic field is assumed to scale with the local
thermal electron density as \citep{Murgia2004,Adam2020} :
\begin{equation}
    B(r) = B_0 \left(\frac{n_e(r)}{n_e(r_0)}\right)^{\eta_B},
    \label{eq:bfield}
\end{equation}
\noindent where $B_0 = 5\ \mu\mathrm{G}$ is the field strength
at reference radius $r_0 = 100\ \mathrm{kpc}$ and $\eta_B =
0.5$ is the scaling index \citep{Bonafede2010}. With the best-fit $\beta$-model
density, Eq.~(\ref{eq:bfield}) yields $B \approx 6.0\
\mu\mathrm{G}$ near the cluster center ($r \sim 1\
\mathrm{kpc}$), declining to ${\sim}1.6\ \mu\mathrm{G}$ at
$R_{500}$ and ${\sim}0.85\ \mu\mathrm{G}$ at the truncation
radius. The radial profile is shown in combined
Figure~\ref{fig:combined_profiles}.

\subsubsection{Cosmic-Ray Spatial Distribution}
\label{sec:cr_spatial}
The number density of both CRp and CRe is scaled with the
thermal electron density as:

\begin{equation}
    n_{\rm CR}(r) \propto n_e(r)^{\eta_{\rm CR}},
    \label{eq:cr_spatial}
\end{equation}

\noindent with $\eta_{\rm CR} = 1$ (isodens scaling) adopted
as the baseline for both species \citep{Brunetti2017, Adam2020}.

\begin{figure}[ht]
    \centering
    \includegraphics[width=0.8\columnwidth]{%
        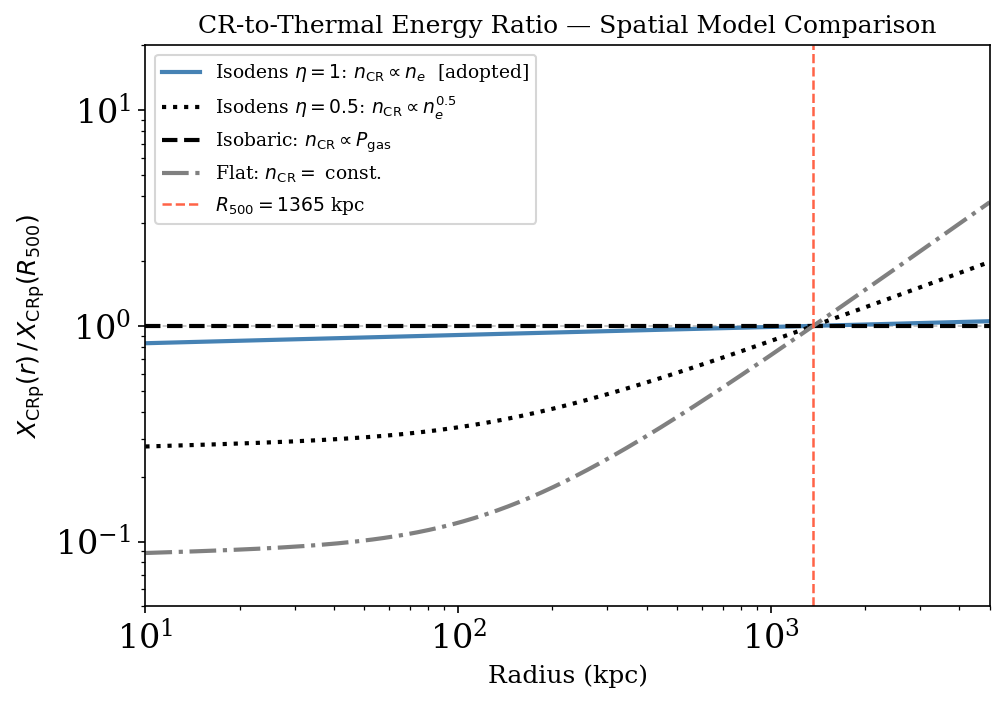}
    \caption{Normalized CR-to-thermal energy ratio
    $X_{\rm CRp}(r)/X_{\rm CRp}(R_{500})$ under four spatial
    models: isodens $\eta=1$ (solid blue, adopted baseline),
    isodens $\eta=0.5$ (dotted), isobaric (dashed), and flat
    (dot-dashed grey). All curves are normalized to unity at
    $R_{500}$ (red dashed line). The four models differ
    by less than $10\%$ within the data range; they diverge
    significantly only in the extrapolated regime.}
    \label{fig:Xcrp_spatial}
\end{figure}

Figure~\ref{fig:Xcrp_spatial} compares four
spatial models, each  normalized at $R_{500}$: isodens $\eta=1$
(baseline), isodens $\eta=0.5$ (shallower concentration),
isobaric ($n_{\rm CR} \propto P_{\rm gas}$), and a flat
(spatially uniform) profile. The isodens and isobaric models
differ by less than $10\%$ within the ACCEPT data range
($r \lesssim 0.22\,R_{500}$), owing to the nearly isothermal
temperature profile ($\alpha = 0.039$). The flat model
diverges most strongly, with $X_{\rm CR}(r)/X_{\rm CR}
(R_{500})$ rising by a factor of ${\sim}4$ at $R_{\rm trunc}$
relative to the baseline.

The CR content is normalized to $R_{500}$ via:
\begin{equation}
    X_{\rm CRp}(R_{500}) =
    \frac{\varepsilon_{\rm CRp}}{\varepsilon_{\rm th}}
    \bigg|_{R_{500}} = 10^{-2} ,
    \qquad
    X_{\rm CRe}(R_{500}) =
    \frac{\varepsilon_{\rm CRe}}{\varepsilon_{\rm th}}
    \bigg|_{R_{500}} = 10^{-5}.
    \label{eq:Xcr_norm}
\end{equation}
These are the fiducial starting values; the sensitivity
of predicted observables to $X_{\rm CRp}$ and $X_{\rm CRe}$
is explored in Section~\ref{sec:results}.

\subsubsection{Cosmic-Ray Energy Spectra}
\label{sec:cr_spectra}

\paragraph{Cosmic-ray protons.}
CRp are modeled with a power-law momentum spectrum:
\begin{equation}
    \frac{\mathrm{d}N_{\rm CRp}}{\mathrm{d}p} \propto
    E^{-\alpha_p}, \qquad \alpha_p = 2.4 .
    \label{eq:crp_spectrum}
\end{equation}
The index $\alpha_p = 2.4$ is consistent with diffusive shock
acceleration at merger-driven shocks
\citep[$\alpha_p \approx 2$--$2.5$;][]{Ryu2003, Kang2007}
and with upper limits from the non-detection of cluster
$\gamma$-ray emission by Fermi-LAT \citep{Ackermann2014}. The proton energy range adopted in the calculations is
$E_p \in [1.22~\mathrm{GeV},\,100~\mathrm{TeV}]$, where the lower
bound corresponds to the kinematic threshold for neutral-pion
production in proton--proton interactions. The upper bound is set to
100~TeV which is adopted as a
representative, physically reasonable upper limit for the proton
distribution, consistent with the pion-decay rule of thumb whereby
protons of energy $E_p$ produce photons with characteristic energy
$E_\gamma \sim E_p/10$; a 100~TeV proton cutoff therefore corresponds
to a $\gamma$-ray cutoff near 10~TeV, above the energy range relevant
to the \textit{Fermi}-LAT comparison presented here. The results and
conclusions of this work are insensitive to the precise value adopted
for the proton maximum energy, as no evidence for a cutoff is present
in the available $\gamma$-ray data.

\paragraph{Primary cosmic-ray electrons.}
Primary CRe are injected with a power-law energy spectrum,
\begin{equation}
    \frac{\mathrm{d}N_{\rm CRe}}{\mathrm{d}E}
    \propto E^{-\alpha_e}, \qquad \alpha_e = 2.4 ,
    \label{eq:cre_spectrum}
\end{equation}
consistent with the integrated spectral index of the A3667
radio relics \citep{JohnstonHollitt2008,Hindson2014}. The injected
population is subsequently evolved using the steady-state energy-loss
treatment implemented in \textsc{MINOT}, which self-consistently
accounts for radiative cooling and yields the equilibrium electron
spectrum. The electron energy range adopted in the calculations is
$E_e \in [1~\mathrm{MeV},\,100~\mathrm{TeV}]$, with the same upper bound
as for the protons and for the same reason: it lies well above the
$\gamma$-ray-relevant energy range and so does not affect the results.

\paragraph{Secondary cosmic-ray electrons.}
Secondary CRe from hadronic $pp$ interactions are computed
self-consistently by MINOT using the \texttt{Pythia8} model
\citep{Sjostrand2015, Kafexhiu2014}. 

\begin{figure}[ht]
    \centering
    \includegraphics[width=0.8\columnwidth]{%
        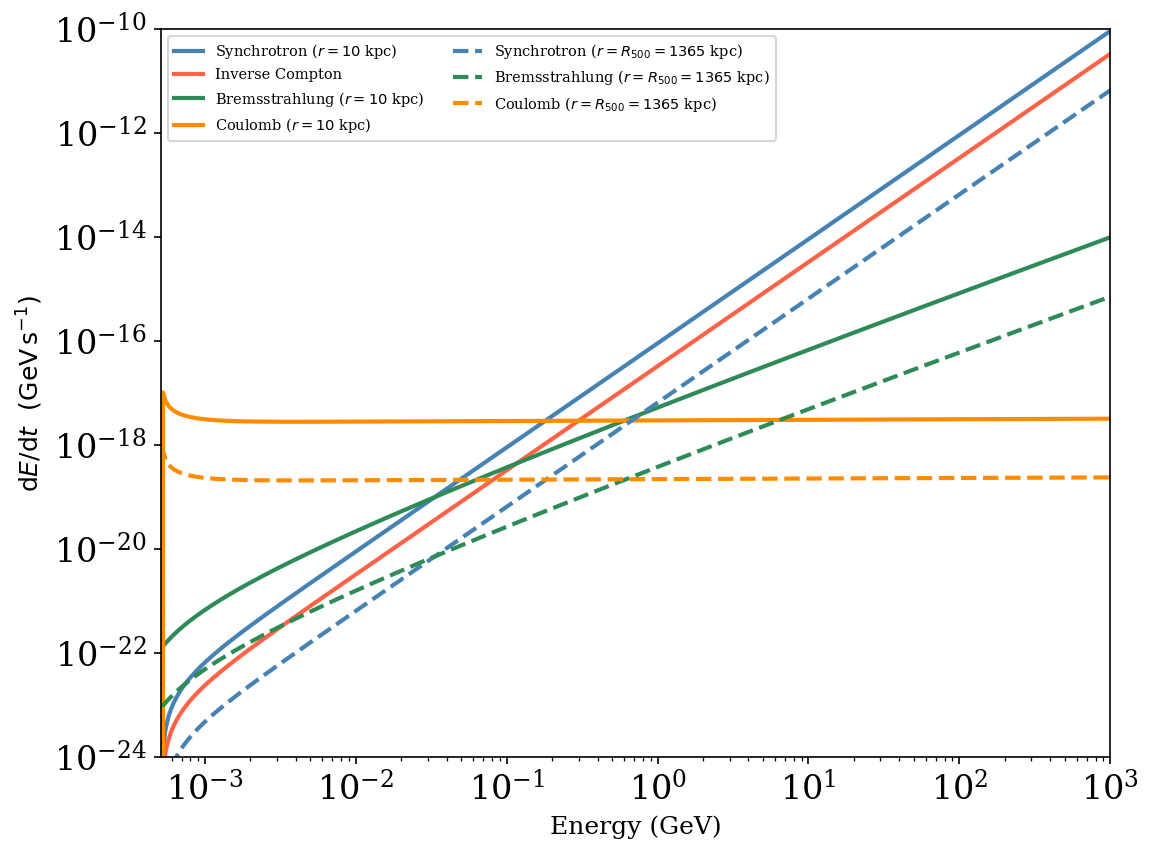}
    \caption{Electron energy loss rates in the A3667 ICM as a function of electron energy, evaluated at $r = 10\
    \mathrm{kpc}$ (solid lines) and $r = R_{500} = 1365\
    \mathrm{kpc}$ (dashed lines). Four loss mechanisms are
    shown: synchrotron (blue), IC (red; single
    curve, identical at all radii as it depends only on the
    CMB radiation field at $z = 0.0556$), bremsstrahlung
    (green), and Coulomb (orange). At $r = 10\ \mathrm{kpc}$,
    $B \approx 6.0\ \mu\mathrm{G} > B_{\rm CMB} \approx
    3.6\ \mu\mathrm{G}$, so synchrotron dominates IC; at
    $r = R_{500}$, $B \approx 1.6\ \mu\mathrm{G} <
    B_{\rm CMB}$ and IC dominates. Coulomb losses control
    the sub-$0.1\ \mathrm{GeV}$ behavior at both radii.}
    \label{fig:eloss}
\end{figure}

\paragraph{Electron energy losses and spectral break.}
The energy loss processes governing the CRe evolution in the
A3667 ICM are illustrated in Figure~\ref{fig:eloss}. Four
mechanisms contribute to $-\mathrm{d}E/\mathrm{d}t$:
synchrotron radiation, IC scattering off
CMB photons, bremsstrahlung, and Coulomb collisions
\citep{Longair2011}. 
Each process is evaluated
at two representative radii — $r = 10\ \mathrm{kpc}$ (inner
core) and $r = R_{500} = 1365\ \mathrm{kpc}$ ($R_{500}$).

At low energies ($E \lesssim 0.1\ \mathrm{GeV}$), Coulomb
losses dominate at both radii, causing electrons to thermalize. The Coulomb rate is proportional to the
local electron density and therefore drops by a factor of
${\sim}14$ between $r = 10\ \mathrm{kpc}$ and $r = R_{500}$,
reflecting the decline in $n_e(r)$ from $4.5 \times 10^{-3}$
to $3.2 \times 10^{-4}\ \mathrm{cm^{-3}}$. At high energies
($E \gtrsim 1\ \mathrm{GeV}$), synchrotron and IC losses
rise steeply as $E^2$, driving efficient radiative cooling.

The relative importance of synchrotron versus IC is set by
the local magnetic energy density compared to that of the CMB.
The equivalent magnetic field of the CMB at $z = 0.0556$ is
$B_{\rm CMB} = \sqrt{8\pi\,U_{\rm CMB}} \approx 3.6\
\mu\mathrm{G}$; synchrotron exceeds IC wherever $B(r) >
B_{\rm CMB}$. With the adopted field profile
(Eq.~\ref{eq:bfield}), this condition holds for $r \lesssim
250\ \mathrm{kpc}$ (${\approx}0.18\,R_{500}$). At the inner
evaluation radius ($r = 10\ \mathrm{kpc}$), $B \approx
6.0\ \mu\mathrm{G} > B_{\rm CMB}$ and synchrotron dominates;
at $r = R_{500}$, $B \approx 1.6\ \mu\mathrm{G} < B_{\rm CMB}$
and IC losses take over. The two evaluation radii therefore bracket the transition between synchrotron-radiation-dominated and IC-dominated cooling, efficiently illustrating both electron cooling regimes within a single figure.

\begin{figure}[ht]
    \centering
    \includegraphics[width=0.9\columnwidth]{%
        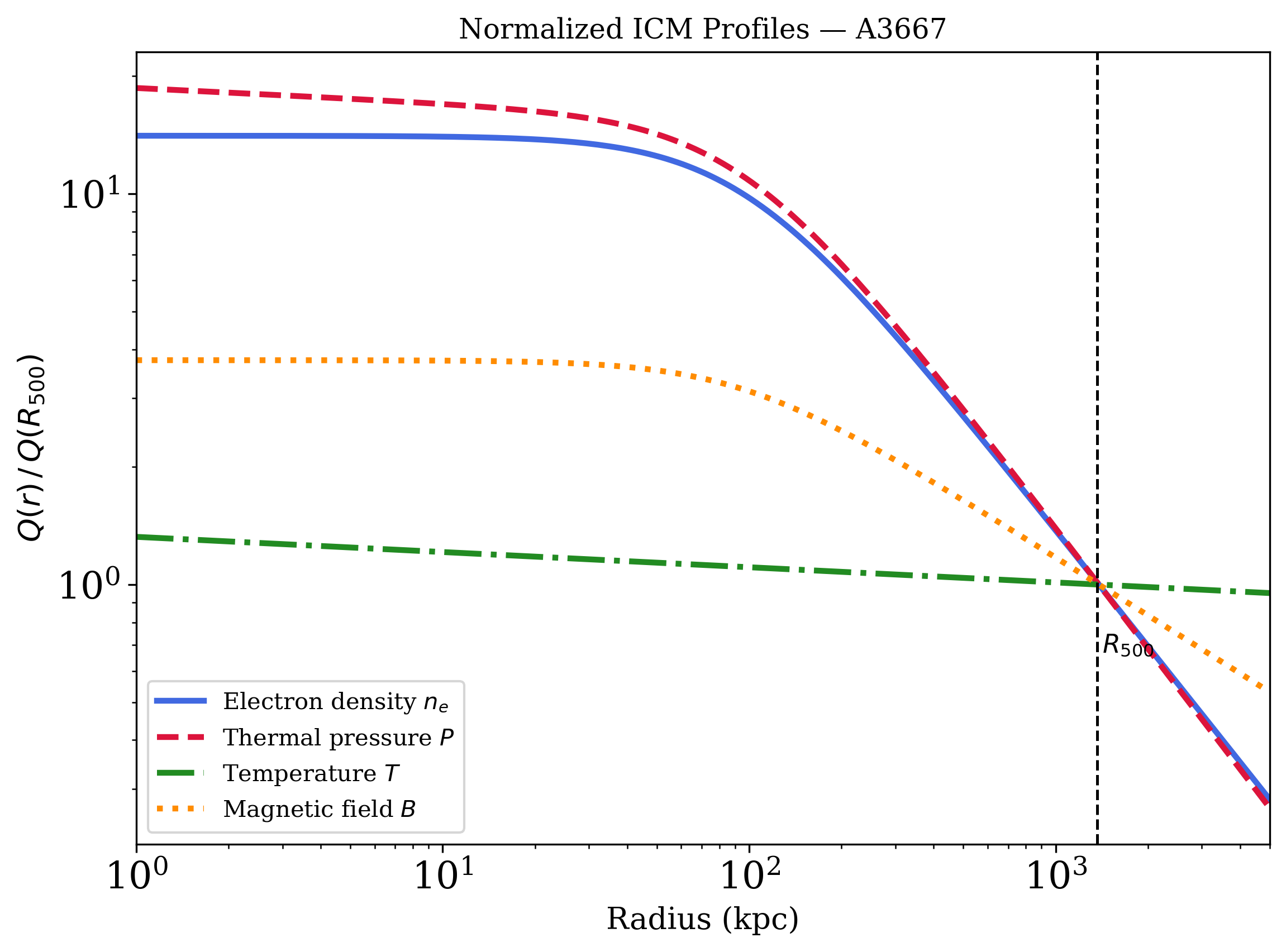}
    \caption{Normalized radial profiles of the four key ICM
    quantities for A3667, each divided by its value at
    $R_{500} = 1365.5\ \mathrm{kpc}$ (vertical black dashed line)
    to facilitate shape comparison on a common scale. The
    electron density $n_e$ (solid blue), thermal pressure $P$
    (dashed red), temperature $T$ (dot-dashed green), and
    magnetic field $B$ (dotted orange) are all evaluated on
    the same logarithmic radial grid from $1$ to
    $5000\ \mathrm{kpc}$.}
    \label{fig:combined_profiles}
\end{figure}

\subsubsection{Map and Numerical Integration Parameters}
\label{sec:minot_setup}
The cluster is modeled on a $6^{\circ}\times 6^{\circ}$ field
of view centered on $(\alpha,\delta) = (303.1763^{\circ},
-56.8432^{\circ})$, with a spatial resolution of $30\
\mathrm{arcsec}$.  The truncation radius subtends $\theta_{\rm trunc} = 1.245^{\circ}$ and the characteristic cluster radius subtends $\theta_{500} = 0.340^{\circ}$, both comfortably within the field of view.

The volume integrals are evaluated on a logarithmic radial grid
with 50 points per decade between $R_{\rm min} = 1\
\mathrm{kpc}$ and $R_{\rm trunc} = 5000\ \mathrm{kpc}$,
providing adequate sampling over four decades in radius.
EBL absorption of $\gamma$-ray emission is modeled using the models in 
\citet{Dominguez2011}. 

\begin{table}[ht]
\centering
\caption{Non-thermal model parameters adopted for the MINOT
initialization of A3667.}
\label{tab:nonthermal_params}
\begin{tabular}{lcc}
\hline\hline
Parameter & Value & Unit \\
\hline
\multicolumn{3}{l}{\textit{Magnetic field}} \\
$B_0$                    & $5.0$   & $\mu\mathrm{G}$ \\
$r_0$                    & $100$   & kpc \\
$\eta_B$                 & $0.5$   & --- \\
$B(r{\sim}0)$            & ${\approx}6.0$ & $\mu\mathrm{G}$ \\
$B(R_{500})$             & ${\approx}1.6$ & $\mu\mathrm{G}$ \\
$B(R_{\rm trunc})$       & ${\approx}0.85$ & $\mu\mathrm{G}$ \\
\hline
\multicolumn{3}{l}{\textit{Cosmic-ray protons}} \\
Spectrum          & PowerLaw              & --- \\
$\alpha_p$        & $2.4$                 & --- \\
$X_{\rm CRp}$     & $10^{-2}$             & --- \\
$E_{p,\rm min}$   & $1.22$                & GeV \\
$E_{p,\rm max}$   & $100$                  & TeV \\
\hline
\multicolumn{3}{l}{\textit{Cosmic-ray electrons}} \\
Injection spectrum & PowerLaw             & --- \\
$\alpha_e$         & $2.4$                & --- \\
Energy-loss model  & Steady               & --- \\
$X_{\rm CRe}$      & $10^{-5}$            & --- \\
$E_{e,\rm min}$    & $1$                  & MeV \\
$E_{e,\rm max}$    & $100$                 & TeV \\
\hline
\multicolumn{3}{l}{\textit{Spatial distribution (both species)}} \\
Model             & Isodens ($\eta_{\rm CR}=1$) & --- \\
\hline
\multicolumn{3}{l}{\textit{Map and integration}} \\
Field of view     & $6\times 6$           & deg \\
Resolution        & $30$                  & arcsec \\
$\theta_{500}$    & $0.340$               & deg \\
$\theta_{\rm trunc}$ & $1.245$            & deg \\
EBL model         & \citet{Dominguez2011} & --- \\
\hline
\end{tabular}
\end{table}

\section{Results}
\label{sec:results}

\subsection{Thermodynamic Structure of A3667}
\label{sec:thermo_results}
Using the MINOT cluster model constructed in
Section~\ref{sec:data_analysis}, we derive the full
thermodynamic structure of A3667 from $1\ \mathrm{kpc}$ to
the truncation radius $R_{\rm trunc} = 5000\ \mathrm{kpc}
\approx 3.7\,R_{500}$. The results are presented in
Figure~\ref{fig:combined_profiles}.

\subsubsection{Electron Density}
\label{sec:ne_result}
The electron density profile, shown in
Figure~\ref{fig:combined_profiles}, is described by the
best-fit $\beta$-model with parameters $n_0 = 4.49 \times 10^{-3}\ \mathrm{cm^{-3}}$, $r_c = 95.5 \ \mathrm{kpc}$, and $\beta = 0.329$. The profile exhibits a flat, nearly constant core within $r \lesssim 20\ \mathrm{kpc}$, followed by a power-law decline at larger radii. At $R_{500}$ the density has declined to $n_e(R_{500}) = 3.11 \times
10^{-4}\ \mathrm{cm^{-3}}$. The profile reaches $n_e \approx 8 \times 10^{-5}\ \mathrm{cm^{-3}}$ at the truncation radius.

\subsubsection{Thermal Pressure}
\label{sec:pressure_result}
The pressure profile $P(r)=n_e(r)\times T(r)$, shown in Figure~\ref{fig:combined_profiles}, decreases from $P \approx 3.03 \times 10^{-2}\ \mathrm{keV\,cm^{-3}}$ in the cluster core to $P(R_{500}) = 1.59 \times 10^{-3}\ \mathrm{keV\,cm^{-3}}$, reaching $P \approx 4 \times 10^{-4}\ \mathrm{keV\,cm^{-3}}$ at the truncation radius.

\subsubsection{Temperature}
\label{sec:temperature_result}
The temperature profile, shown in
Figure~\ref{fig:combined_profiles}, is nearly isothermal
across the full radial range, consistent with the power-law
fit $T(r) = T_0\,(r/r_0)^{-\alpha}$ with $\alpha = 0.039$.
The temperature declines only modestly from $T \approx
5.9\ \mathrm{keV}$ at $r = 10\ \mathrm{kpc}$ to $T(R_{500})
= 5.09\ \mathrm{keV}$ at $R_{500}$. 

\subsection{Predicted Gamma-ray Emission from Hadronic Interactions}
\label{sec:gamma_results}
We compute the gamma-ray emission from A3667 arising
from hadronic cosmic-ray proton interactions with the thermal
ICM via inelastic proton--proton ($pp$) collisions:

\begin{equation}
    p + p \rightarrow p + p + \pi^0 + \ldots,
    \qquad \pi^0 \rightarrow \gamma + \gamma.
    \label{eq:pp_chain}
\end{equation}

\noindent The resulting gamma-ray flux depends on the spatial
distribution and energy spectrum of the CRp population, the
thermal gas density, and the cluster volume. All calculations
are performed using the MINOT framework with the Pythia8
hadronic interaction model \citep{Adam2020}, using the
baseline cluster model described in
Section~\ref{sec:minot_init}. We additionally compute the
contribution from IC scattering of primary
CRe off CMB photons as a reference channel.

\subsubsection{Hadronic Gamma-ray Spectrum and Parameter Dependence}
\label{sec:gamma_spectrum}

\begin{figure*}[ht]
    \centering
    \includegraphics[width=\textwidth]{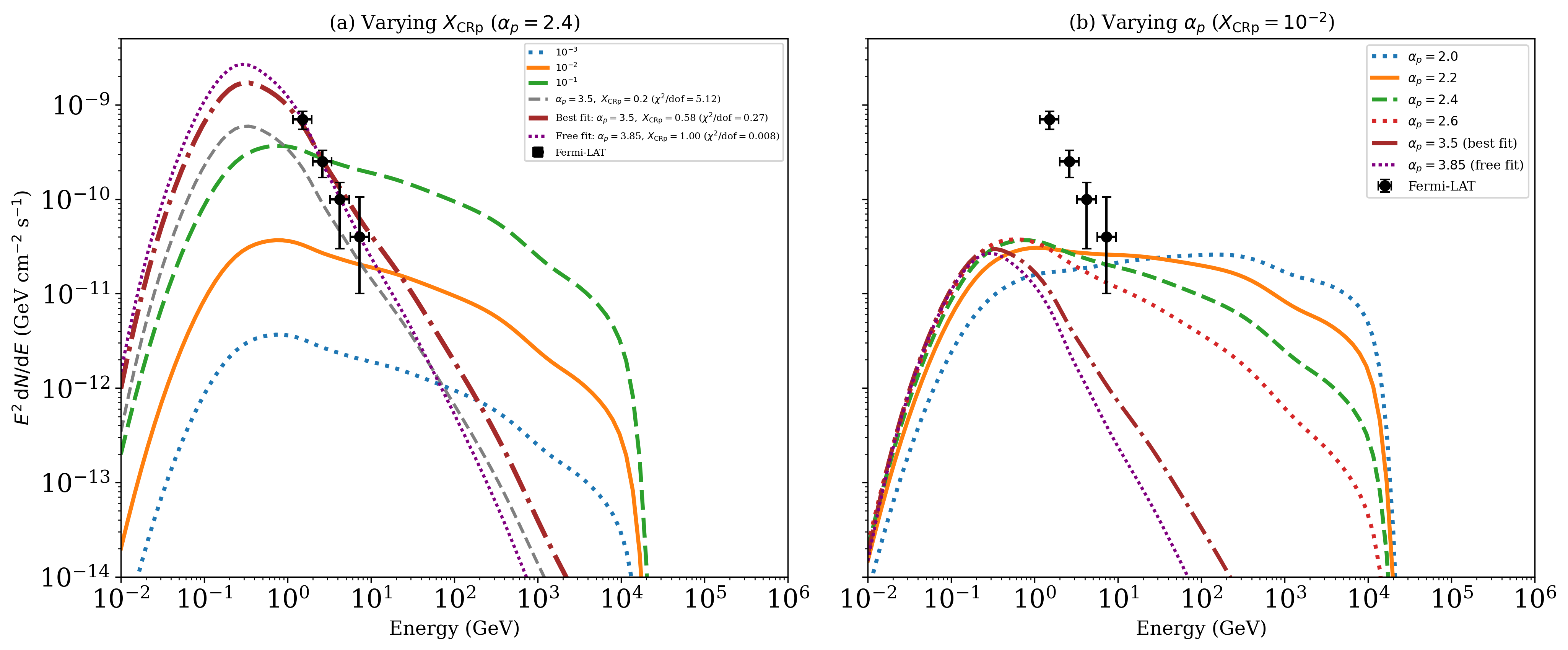}
    \caption{Dependence of the predicted hadronic $\gamma$-ray spectrum on the
adopted cosmic-ray proton parameters. \textbf{Left:} variation of the CR
proton energy fraction $X_{\rm CRp}$ over $10^{-3}$, $10^{-2}$, and
$10^{-1}$ while keeping the proton spectral index fixed at $\alpha_p=2.4$.
Increasing $X_{\rm CRp}$ changes primarily the overall normalization of the
spectrum. The gray dashed curve shows $\alpha_p=3.5$, $X_{\rm CRp}=0.2$, as
originally suggested by the referee. The brown dash-dotted curve shows the
corresponding best-fit solution at fixed $\alpha_p=3.5$, obtained by
fitting $X_{\rm CRp}$ to the \textit{Fermi}-LAT data
($X_{\rm CRp}=0.575\pm0.098$, $\chi^2/{\rm dof}=0.28$), which provides a
substantially better match to the observed spectrum than the referee's
initial guess ($\chi^2/{\rm dof}=5.13$). The purple dotted curve shows the
result of a fully free fit, with both $\alpha_p$ and $X_{\rm CRp}$ left as
free parameters ($\alpha_p\approx3.85$, $X_{\rm CRp}\approx1.0$); with only
four spectral points this two-parameter fit is statistically underconstrained
and converges to an unphysically high CR-to-thermal energy ratio, and we do
not consider it a reliable measurement, but include it for completeness.
\textbf{Right:} variation of the proton spectral index $\alpha_p$ over
2.0, 2.2, 2.4, and 2.6 while fixing $X_{\rm CRp}=10^{-2}$, together with the
best-fit index $\alpha_p=3.5$ (brown dash-dotted) and the free-fit index
$\alpha_p\approx3.85$ (purple dotted) at the same reference
$X_{\rm CRp}=10^{-2}$, shown for direct comparison of spectral shape with
the illustrative curves; note that these two curves use the fitted spectral
indices but not the fitted $X_{\rm CRp}$ values from the left panel. Black
points denote the observed \textit{Fermi}-LAT spectrum of A3667
\citep{Manna2024}. Together, these comparisons demonstrate the sensitivity
of the predicted hadronic emission to both the CR energy density and the
proton spectral slope, and identify a physically plausible, if extreme,
parameter combination (low Mach number shock, high CR-to-thermal
equipartition) consistent with the observed $\gamma$-ray emission.}
    \label{fig:xcrp_dependence}
\end{figure*}

Figure~\ref{fig:xcrp_dependence} shows the predicted hadronic
$\gamma$-ray spectral energy distribution (SED),
$E^2\,\mathrm{d}N/\mathrm{d}E$, integrated over the full cluster
volume out to $R_{\rm trunc}=5000\ \mathrm{kpc}$. The baseline model
adopts a CR proton spectral index $\alpha_p=2.4$, an isodens spatial
distribution ($\eta=1$), and a CR proton energy fraction
$X_{\rm CRp}=10^{-2}$ following the fiducial MINOT configuration
\citep{Adam2020}.  

The left panel of Figure~\ref{fig:xcrp_dependence} shows the effect of
varying the CR proton energy fraction over $X_{\rm CRp}=10^{-3}$,
$10^{-2}$, and $10^{-1}$,  while keeping $\alpha_p=2.4$ fixed.
Increasing $X_{\rm CRp}$ primarily changes the overall normalization
of the $\gamma$-ray spectrum, whereas the spectral shape remains
essentially unchanged because it is determined by the underlying
proton energy distribution. We
now explore a solution with a steeper spectral index,
$\alpha_p=3.5$, corresponding to a shock Mach number $M=1.9$ via the
diffusive shock acceleration relation between compression ratio and
Mach number~\cite{Zimbardo}, together with a correspondingly higher CR proton energy
fraction. Fixing $\alpha_p=3.5$ and fitting $X_{\rm CRp}$ to the four
\textit{Fermi}-LAT spectral points via a least-squares
($\chi^2$-minimization) fit, we obtain a best-fit value
$X_{\rm CRp}=0.575\pm0.098$, with $\chi^2/{\rm dof}=0.28$
($\mathrm{dof}=3$), which reproduces the observed spectrum
substantially better than 
$X_{\rm CRp}=0.2$, for which $\chi^2/{\rm dof}=5.13$. We also tested a
fully free fit in which both $\alpha_p$ and $X_{\rm CRp}$ are left as
free parameters with only four spectral data points. This
two-parameter fit converges to
$\alpha_p\approx3.85$, $X_{\rm CRp}\approx1.0$, i.e.\ a cosmic-ray
proton energy density comparable to (in equipartition with) the
thermal energy density of the cluster, as is observed locally in the
Milky Way~\cite{Ferriere2001}. We therefore adopt the fixed-$\alpha_p=3.5$
fit as our fiducial best-fit solution, while showing the fully free
fit in Figure~\ref{fig:xcrp_dependence} for completeness.

The right panel of Figure~\ref{fig:xcrp_dependence}  illustrates the dependence on the proton spectral
index for $\alpha_p=2.0$, $2.2$, $2.4$, and $2.6$ while fixing
$X_{\rm CRp}=10^{-2}$, together with the best-fit and free-fit
spectral indices, $\alpha_p=3.5$ and $\alpha_p\approx3.85$
respectively, shown at the same reference $X_{\rm CRp}=10^{-2}$ for
direct comparison of spectral shape.

Comparison with the observed \textit{Fermi}-LAT spectral measurements
of A3667 demonstrates that both the CR proton energy density and the
proton spectral index strongly influence the predicted hadronic
emission. The fixed-$\alpha_p=3.5$ best-fit solution combined with a relatively high
CR-to-thermal equipartition fraction ($X_{\rm CRp}\sim0.6$), which is
on the extreme end of the parameter space explored in this work. However, $X_{\rm CRp}$ parametrizes
energy equipartition between cosmic rays and the thermal gas rather
than shock acceleration efficiency directly, and values of this order
are not without precedent, being comparable to the local
CR-to-thermal energy density ratio in the Milky Way~\cite{Ferriere2001}. We therefore
regard this solution as extreme but not excluded, and note that it
substantially improves consistency with the observed $\gamma$-ray
spectrum relative to the fiducial parameters. Accordingly, the adopted baseline model
values $X_{\rm CRp}=10^{-2}$ and $\alpha_p=2.4$~\ref{sec:minot_setup} should be regarded as
fiducial reference parameters rather than uniquely determined physical
quantities, with $\alpha_p=3.5$, $X_{\rm CRp}\approx0.6$ representing
a statistically preferred, if physically extreme, alternative.

\subsubsection{Enclosed Gamma-ray Flux}
\label{sec:gamma_flux}

\begin{figure}[ht]
    \centering
    \includegraphics[width=0.8\columnwidth]{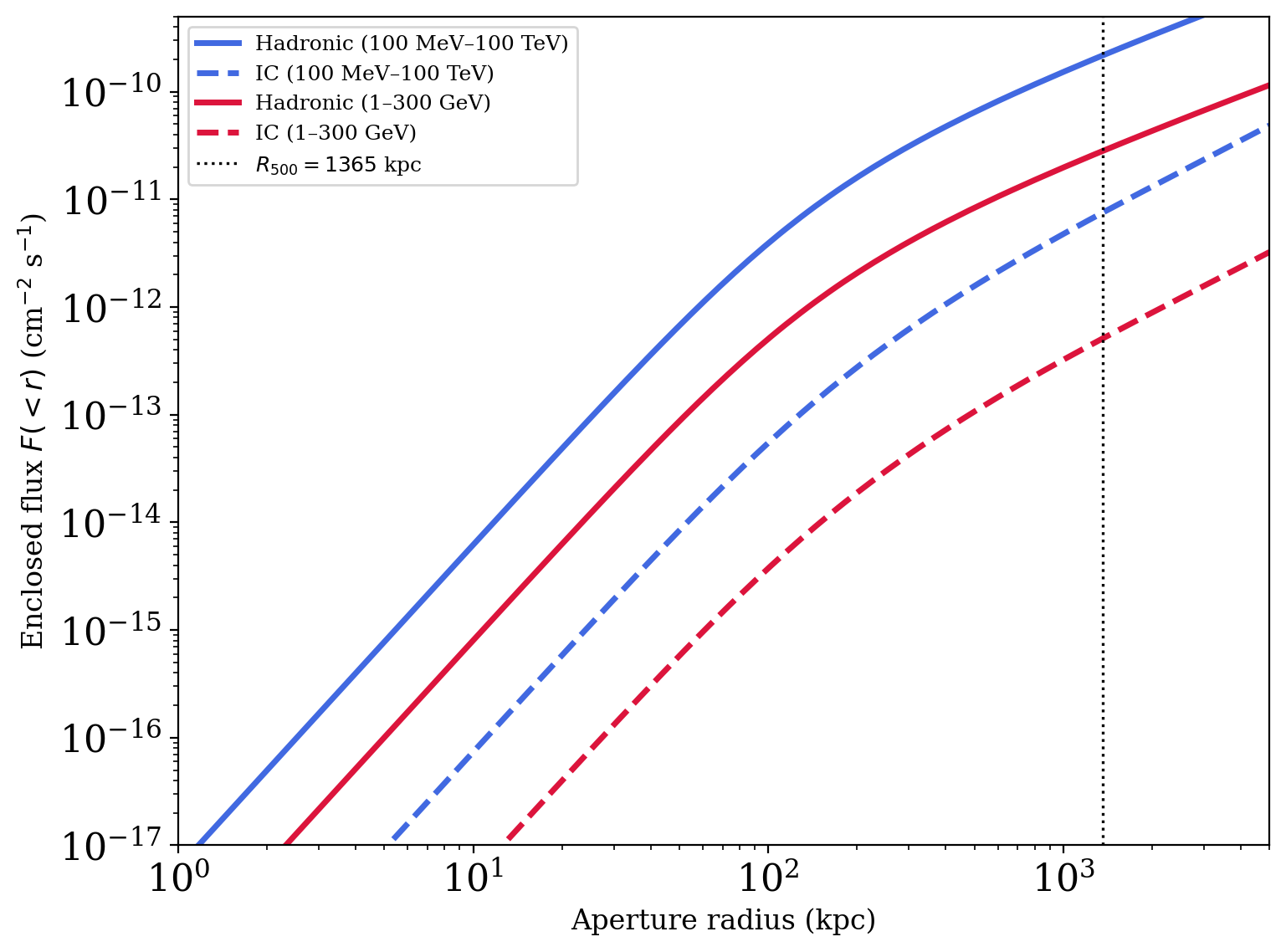}
    \caption{Enclosed non-thermal $\gamma$-ray flux $F(<r)$ as a
    function of aperture radius for A3667, comparing the hadronic
    ($pp$ interactions; solid lines) and IC (dashed lines) components. Blue curves correspond to the broad
    $100~\mathrm{MeV}$--$100~\mathrm{TeV}$ energy range, while red
    curves show the $1$--$300~\mathrm{GeV}$ band adopted in our
    Fermi-LAT analysis. In both energy intervals, the hadronic
    component dominates over the IC emission by more than an order
    of magnitude. The predicted hadronic enclosed flux within
    $R_{\rm trunc}$ is
    $8.89\times10^{-10}\ \mathrm{cm^{-2}\,s^{-1}}$
    ($100~\mathrm{MeV}$--$100~\mathrm{TeV}$) and
    $1.15\times10^{-10}\ \mathrm{cm^{-2}\,s^{-1}}$
    ($1$--$300~\mathrm{GeV}$), the latter being comparable to the
    Fermi-LAT flux of
    $1.3\times10^{-10}\ \mathrm{cm^{-2}\,s^{-1}}$
    reported by \citet{Manna2024}. The vertical dashed line
    marks $R_{500}$.}
    \label{fig:gamma_flux}
\end{figure}

Figure~\ref{fig:gamma_flux} shows the enclosed hadronic and
IC $\gamma$-ray fluxes as a function of the aperture
radius for A3667. The enclosed flux increases monotonically with
radius, reflecting the extended distribution of non-thermal emission
throughout the intracluster medium. For the baseline hadronic model
($X_{\rm CRp}=10^{-2}$, $\alpha_p=2.4$, $\eta=1$), only
$\sim24\%$ of the total hadronic emission originates within
$R_{500}$, while the remaining $\sim76\%$ is produced between
$R_{500}$ and the adopted truncation radius,
$R_{\rm trunc}=5000\ \mathrm{kpc}$.

The predicted enclosed hadronic fluxes for the baseline model and the
various CR parameter choices are summarized in
Table~\ref{tab:gamma_flux_variation}. For the baseline model, the
integrated flux in the broad
$100~\mathrm{MeV}$--$100~\mathrm{TeV}$ band increases from
$2.18\times10^{-10}\ \mathrm{cm^{-2}\,s^{-1}}$ within $R_{500}$ to
$8.89\times10^{-10}\ \mathrm{cm^{-2}\,s^{-1}}$ within
$R_{\rm trunc}$. In the
$1$--$300~\mathrm{GeV}$ band corresponding to our Fermi-LAT analysis,
the enclosed flux increases from
$2.82\times10^{-11}\ \mathrm{cm^{-2}\,s^{-1}}$ to
$1.15\times10^{-10}\ \mathrm{cm^{-2}\,s^{-1}}$, in good agreement
with the Fermi-LAT flux of
$1.3\times10^{-10}\ \mathrm{cm^{-2}\,s^{-1}}$~\cite{Manna2024}.

The integrated flux exhibits a nearly linear dependence on the adopted
CR proton energy fraction, increasing by two orders of magnitude as
$X_{\rm CRp}$ is varied from $10^{-3}$ to $10^{-1}$.
In contrast, changing the proton spectral index primarily
redistributes the emitted power with energy. In the
$1$--$300~\mathrm{GeV}$ band, the enclosed flux decreases from
$1.15\times10^{-10}\ \mathrm{cm^{-2}\,s^{-1}}$ for
$\alpha_p=2.2$--2.4 to
$5.89\times10^{-11}\ \mathrm{cm^{-2}\,s^{-1}}$ for
$\alpha_p=3.0$, while a harder spectrum
($\alpha_p=2.0$) yields
$7.32\times10^{-11}\ \mathrm{cm^{-2}\,s^{-1}}$. The broad-band
($100~\mathrm{MeV}$--$100~\mathrm{TeV}$) flux is less sensitive to
the spectral index because the total hadronic emission is redistributed
over a wider energy interval. Finally, adopting a shallower CR spatial
distribution ($n_{\rm CRp}\propto n_e^{0.5}$) slightly decreases the
flux within $R_{500}$ (from
$2.18\times10^{-10}$ to
$1.94\times10^{-10}\ \mathrm{cm^{-2}\,s^{-1}}$) but increases the
total enclosed flux within $R_{\rm trunc}$ by approximately
$60\%$, owing to the larger CR proton content in the cluster
outskirts. These results demonstrate that the predicted hadronic
emission is most strongly governed by the CR proton energy density,
while the proton spectral index controls the energy distribution of
the emitted $\gamma$ rays and the CR spatial profile primarily
determines how the emission is distributed throughout the cluster.

For comparison, we also investigated whether the observed
Fermi-LAT integrated flux can be reproduced by adopting the measured
photon spectral index ($\Gamma = 3.61 \pm 0.33$) as an illustrative
spectral slope in the hadronic model. For the fiducial normalization
($X_{\rm CRp}=10^{-2}$), the predicted
$1$--$300~\mathrm{GeV}$ flux within $R_{\rm trunc}$ is
$2.67\times10^{-11}\ \mathrm{cm^{-2}\,s^{-1}}$, approximately a
factor of five below the observed value of
$1.3\times10^{-10}\ \mathrm{cm^{-2}\,s^{-1}}$~\cite{Manna2024}. Since the hadronic $\gamma$-ray
flux scales nearly linearly with the CR proton energy fraction,
matching the observed integrated flux requires
$X_{\rm CRp}\approx4.95\times10^{-2}$. With this normalization, the
predicted enclosed flux becomes
$1.32\times10^{-10}\ \mathrm{cm^{-2}\,s^{-1}}$ within
$R_{\rm trunc}$, in excellent agreement with the Fermi-LAT
measurement. This comparison demonstrates that, for a spectrum as
steep as the observed Fermi-LAT photon index, a substantially larger
CR proton energy density is required than the fiducial value adopted
in the baseline model. More generally, it highlights that the CR
proton energy fraction and spectral slope must be constrained
simultaneously when interpreting the $\gamma$-ray emission from
A3667, rather than assuming a fixed fiducial normalization.

\begin{table*}[ht]
\centering
\caption{Predicted enclosed hadronic $\gamma$-ray fluxes for A3667
under different CR proton model assumptions. Fluxes are computed using
spherical integration within $R_{500}$ and the truncation radius
($R_{\rm trunc}=5000$ kpc). The final row gives an observational comparison in which the CR proton normalization is adjusted to reproduce the observed Fermi-LAT
$1$--$300~\mathrm{GeV}$ integrated flux for the measured photon
spectral index ($\Gamma=3.61$).}
\label{tab:gamma_flux_variation}
\small
\begin{tabular}{lccccc}
\hline\hline
Model &
Parameter &
\multicolumn{2}{c}{$100~\mathrm{MeV}$--$100~\mathrm{TeV}$} &
\multicolumn{2}{c}{$1$--$300~\mathrm{GeV}$} \\
\cline{3-6}
&& $R_{500}$ & $R_{\rm trunc}$ &
$R_{500}$ & $R_{\rm trunc}$ \\
\hline
Baseline
&
$\alpha_p=2.4,\ \eta=1,\ X_{\rm CRp}=10^{-2}$
&
$2.18\times10^{-10}$
&
$8.89\times10^{-10}$
&
$2.82\times10^{-11}$
&
$1.15\times10^{-10}$
\\
\hline
\multicolumn{6}{l}{\textit{Variation of CR proton energy fraction}}\\
$X_{\rm CRp}=10^{-3}$
&
&
$2.18\times10^{-11}$
&
$8.89\times10^{-11}$
&
$2.82\times10^{-12}$
&
$1.15\times10^{-11}$
\\
$5\times10^{-3}$
&
&
$1.09\times10^{-10}$
&
$4.45\times10^{-10}$
&
$1.41\times10^{-11}$
&
$5.76\times10^{-11}$
\\
$10^{-2}$
&
&
$2.18\times10^{-10}$
&
$8.89\times10^{-10}$
&
$2.82\times10^{-11}$
&
$1.15\times10^{-10}$
\\
$3\times10^{-2}$
&
&
$6.54\times10^{-10}$
&
$2.67\times10^{-9}$
&
$8.47\times10^{-11}$
&
$3.45\times10^{-10}$
\\
$10^{-1}$
&
&
$2.18\times10^{-9}$
&
$8.89\times10^{-9}$
&
$2.82\times10^{-10}$
&
$1.15\times10^{-9}$
\\
\hline
\multicolumn{6}{l}{\textit{Variation of proton spectral index}}\\
$\alpha_p=2.0$
&
&
$7.82\times10^{-11}$
&
$3.19\times10^{-10}$
&
$1.79\times10^{-11}$
&
$7.32\times10^{-11}$
\\
$\alpha_p=2.2$
&
&
$1.65\times10^{-10}$
&
$6.74\times10^{-10}$
&
$2.82\times10^{-11}$
&
$1.15\times10^{-10}$
\\
$\alpha_p=2.4$
&
&
$2.18\times10^{-10}$
&
$8.89\times10^{-10}$
&
$2.82\times10^{-11}$
&
$1.15\times10^{-10}$
\\
$\alpha_p=2.6$
&
&
$2.36\times10^{-10}$
&
$9.64\times10^{-10}$
&
$2.36\times10^{-11}$
&
$9.65\times10^{-11}$
\\
$\alpha_p=2.8$
&
&
$2.37\times10^{-10}$
&
$9.68\times10^{-10}$
&
$1.87\times10^{-11}$
&
$7.62\times10^{-11}$
\\
$\alpha_p=3.0$
&
&
$2.30\times10^{-10}$
&
$9.40\times10^{-10}$
&
$1.44\times10^{-11}$
&
$5.89\times10^{-11}$
\\
\hline
\multicolumn{6}{l}{\textit{Variation of CR spatial distribution}}\\
$\eta=1.0$
&
Isodens
&
$2.18\times10^{-10}$
&
$8.89\times10^{-10}$
&
$2.82\times10^{-11}$
&
$1.15\times10^{-10}$
\\
$\eta=0.5$
&
$n_{\rm CRp}\propto n_e^{0.5}$
&
$1.94\times10^{-10}$
&
$1.44\times10^{-9}$
&
$2.52\times10^{-11}$
&
$1.86\times10^{-10}$
\\
\hline
\hline
\multicolumn{6}{l}{\textit{Comparison with the observed Fermi-LAT spectrum}}\\
$\Gamma=3.61,\;
X_{\rm CRp}=4.95\times10^{-2}$
&
&
$9.68\times10^{-10}$
&
$3.95\times10^{-9}$
&
$3.25\times10^{-11}$
&
$1.32\times10^{-10}$
\\
\hline\hline
\multicolumn{6}{l}{\textit{Referee-requested best-fit solutions ($\alpha_p=3.5$ regime)}}\\
Referee suggested
&
$\alpha_p=3.5,\ X_{\rm CRp}=0.2$
&
$4.04\times10^{-9}$
&
$1.65\times10^{-8}$
&
$1.51\times10^{-10}$
&
$6.15\times10^{-10}$
\\
Best fit (fixed $\alpha_p$)
&
$\alpha_p=3.5,\ X_{\rm CRp}=0.575$
&
$1.16\times10^{-8}$
&
$4.74\times10^{-8}$
&
$4.33\times10^{-10}$
&
$1.77\times10^{-9}$
\\
Fully free fit
&
$\alpha_p=3.85,\ X_{\rm CRp}\approx1.00$
&
$1.82\times10^{-8}$
&
$7.44\times10^{-8}$
&
$4.89\times10^{-10}$
&
$1.99\times10^{-9}$
\\
\hline\hline
\end{tabular}
\end{table*}

\subsubsection{Gamma-ray Surface Brightness Map}
\label{sec:gamma_map}
\begin{figure}[ht]
    \centering
    \includegraphics[width=0.8\columnwidth]{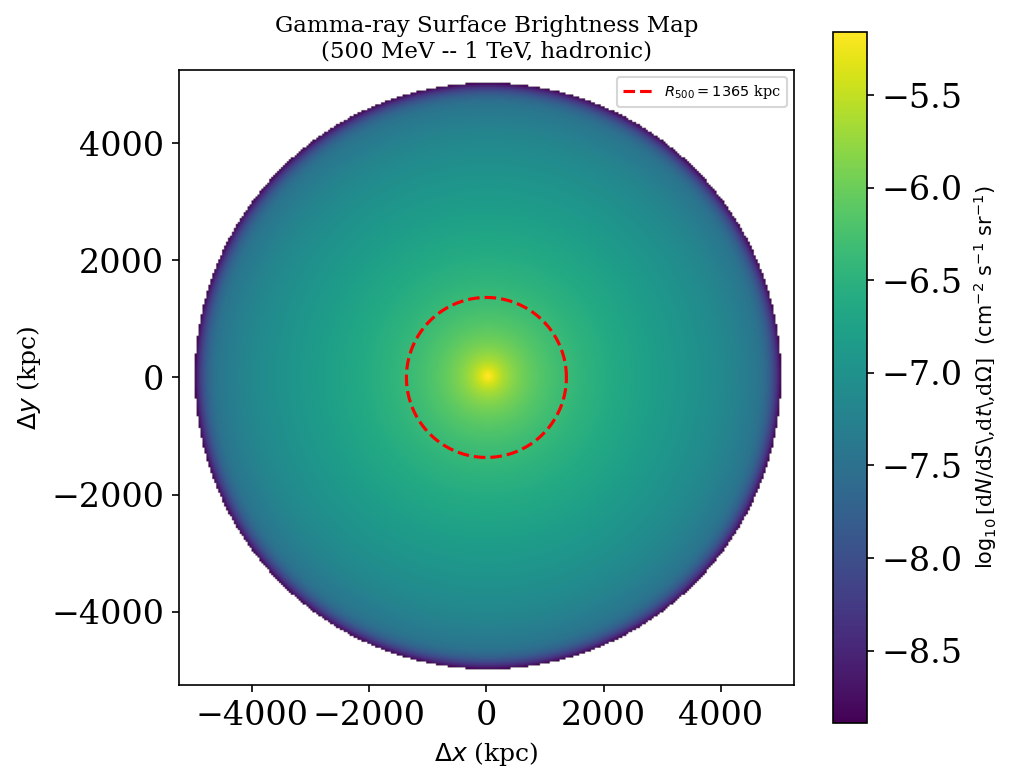}
    \caption{Predicted gamma-ray surface brightness map of
    A3667 in the $500\ \mathrm{MeV}$--$1\ \mathrm{TeV}$
    band from hadronic $pp$ interactions, shown on a
    logarithmic colour scale. The red dashed circle marks
    $R_{500} = 1365\ \mathrm{kpc}$. The map is clipped to $\pm R_{\rm trunc}$ in both axes.
    The circular symmetry reflects the azimuthally averaged
    input profiles; real asymmetries due to the merging
    morphology are not included.}
    \label{fig:gamma_map}
\end{figure}

Figure~\ref{fig:gamma_map} shows the two-dimensional
gamma-ray surface brightness map of A3667 in the $500\
\mathrm{MeV}$--$1\ \mathrm{TeV}$ band. The emission is
circularly symmetric and centrally concentrated, peaking
at the cluster center with a surface brightness of
${\sim}10^{-5.5}\ \mathrm{cm^{-2}\ s^{-1}\ sr^{-1}}$
and declining to ${\sim}10^{-8.5}\ \mathrm{cm^{-2}\
s^{-1}\ sr^{-1}}$ at the cluster outskirts, spanning a
dynamic range of three orders of magnitude. The circular
symmetry reflects the azimuthally averaged nature of the
input thermodynamic profiles; in reality, the merging
morphology of A3667  with its prominent cold fronts and
radio relics would produce significant asymmetries in
the actual CR and gas distributions not captured by
our spherically symmetric model.

The red dashed circle marks $R_{500} = 1365\ \mathrm{kpc}$,
corresponding to an angular radius of $\theta_{500} =
0.340^{\circ}$ at the cluster redshift. The brightest emission is
strongly concentrated within $R_{500}$, although as shown
in Figure~\ref{fig:gamma_flux} the integrated flux is
dominated by the extended emission beyond this radius.

\subsection{Predicted IC Emission from Cosmic-Ray Electrons}
\label{sec:ic_results}
We compute the predicted gamma-ray emission from A3667 arising
from IC scattering of cosmic-ray electrons
(CRe) off the cosmic microwave background (CMB) radiation
field. We separately treat two CRe populations:
\textit{primary} electrons injected directly by acceleration
processes at merger shocks and turbulence, and
\textit{secondary} electrons produced as decay products of
charged pions from hadronic $pp$ interactions. All
calculations are performed using the MINOT framework with the
baseline cluster model described in
Section~\ref{sec:minot_init}, using an energy grid spanning
$10^{-6}$--$10^{4}$\,GeV over the radial range
$1$--$5000$\,kpc. We additionally compute the total IC
emission (primary plus secondary) across three energy bands
and compare with the hadronic pion-decay predictions of
Section~\ref{sec:gamma_results}.

\subsubsection{Inverse-Compton Spectrum and Parameter Dependence}
\label{sec:ic_spectrum}

\begin{figure*}[ht]
    \centering
    \includegraphics[width=\textwidth]{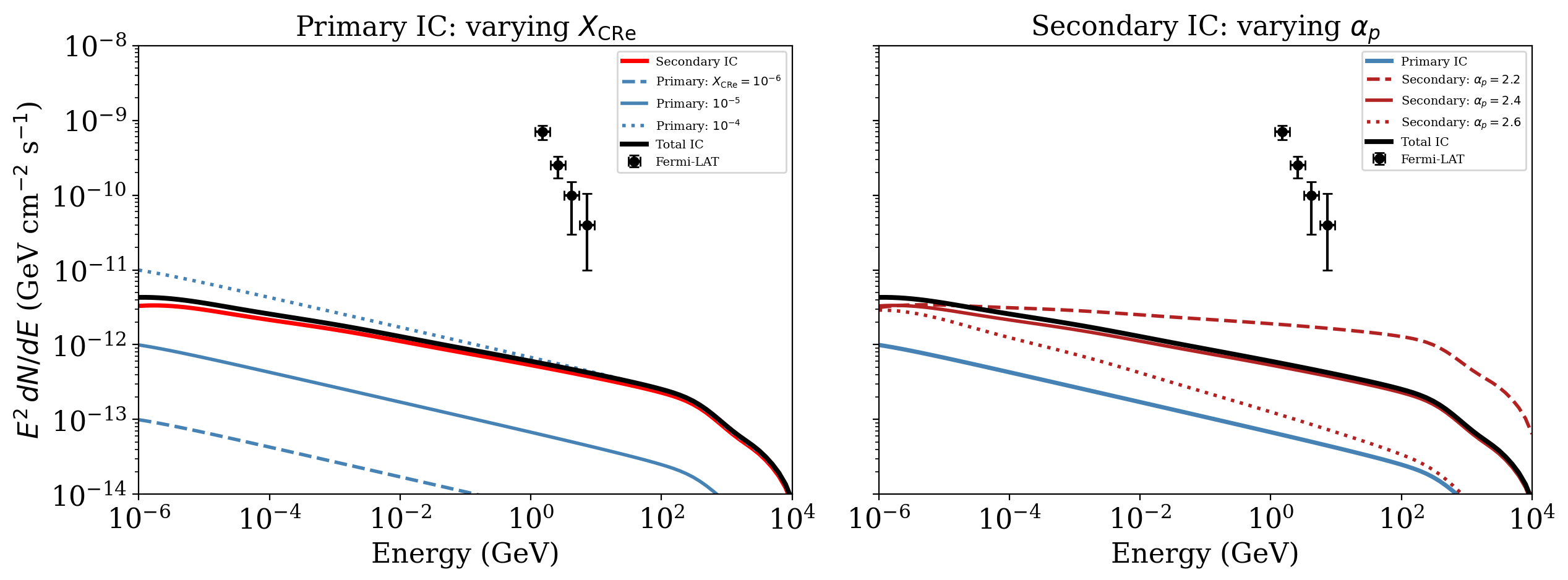}
    \caption{Dependence of the predicted IC spectrum on the
    assumed cosmic-ray populations. \textbf{Left:} Primary IC emission
    for three values of the primary cosmic-ray electron energy fraction,
    $X_{\rm CRe}=10^{-6}$, $10^{-5}$ (baseline), and $10^{-4}$, while
    the secondary IC component is kept fixed. Increasing
    $X_{\rm CRe}$ changes only the normalization of the primary IC
    emission, whereas the spectral shape remains unchanged.
    \textbf{Right:} Secondary IC emission for three proton spectral
    indices, $\alpha_p=2.2$, $2.4$, and $2.6$, while the primary IC
    component is held fixed. Because the secondary electrons are
    produced through hadronic interactions, their IC spectrum directly
    reflects the spectral properties of the parent CR proton
    population. In both panels the black curve denotes the total IC
    emission predicted by the baseline model, and the black points show
    the observed \textit{Fermi}-LAT spectral measurements of A3667
    \citep{Manna2024}.}
    \label{fig:ic_dependence}
\end{figure*}

Figure~\ref{fig:ic_dependence} shows the predicted IC spectral energy distribution (SED),
$E^2\,\mathrm{d}N/\mathrm{d}E$, integrated over the full cluster
volume out to $R_{\rm trunc}=5000~\mathrm{kpc}$. The IC emission is
produced by both primary cosmic-ray electrons (CRe), injected directly
into the intracluster medium, and secondary CRe generated through
hadronic $pp$ interactions involving CR protons. Throughout this work,
the primary electron population is evolved using the steady-state
cooling treatment implemented in \textsc{MINOT}, so that the predicted
IC spectrum is computed from the equilibrium electron distribution
rather than the injected spectrum.

The left panel of Figure~\ref{fig:ic_dependence} illustrates the
dependence of the IC spectrum on the primary CR electron energy
fraction. Three values,
$X_{\rm CRe}=10^{-6}$, $10^{-5}$ (baseline), and
$10^{-4}$, are considered,  while all other model parameters are kept
fixed. Increasing $X_{\rm CRe}$ changes only the normalization of the
primary IC emission, whereas its spectral shape remains essentially
unchanged because the injected electron spectral index is fixed.
Throughout the energy range considered, the secondary IC component is
also shown for comparison, together with the total IC emission
predicted by the baseline model.

The right panel shows the dependence of the secondary IC emission on
the parent CR proton spectrum. Three proton spectral indices,
$\alpha_p=2.2$, $2.4$, and $2.6$, are considered while fixing
$X_{\rm CRp}=10^{-2}$. Since the secondary electrons are produced
self-consistently through hadronic interactions, their equilibrium
energy distribution inherits the spectral properties of the parent CR
protons. Consequently, changing $\alpha_p$ modifies both the
normalization and the spectral slope of the secondary IC component,
whereas the primary IC emission remains unchanged. At photon energies
above a few hundred GeV, both the primary and secondary IC spectra
exhibit a pronounced steepening caused by radiative cooling of the
highest-energy electrons through synchrotron and IC 
losses, which are treated self-consistently in the equilibrium
electron spectrum calculated by \textsc{MINOT}.

For the fiducial model
($X_{\rm CRe}=10^{-5}$,
$X_{\rm CRp}=10^{-2}$,
$\alpha_p=2.4$),
the total IC emission remains more than an order of magnitude below
the corresponding hadronic $\gamma$-ray emission over the
$1$--$300~\mathrm{GeV}$ energy range considered in our Fermi-LAT
analysis. The parameter study demonstrates that the normalization of
the primary IC component is governed primarily by the CR electron
energy density, whereas the spectral properties of the secondary IC
component are controlled by the parent CR proton population. The
adopted values of $X_{\rm CRe}=10^{-5}$ and
$\alpha_p=2.4$ should therefore be regarded as fiducial model
parameters rather than uniquely determined physical quantities.

\subsubsection{Enclosed Inverse-Compton Flux}
\label{sec:ic_flux}

Figure~\ref{fig:gamma_flux} compares the enclosed hadronic and
IC $\gamma$-ray fluxes as a function of aperture
radius for A3667. As in the hadronic case, the enclosed IC flux
increases monotonically with aperture radius, reflecting the extended
distribution of relativistic electrons throughout the intracluster
medium. For the fiducial model, the total IC emission remains
subdominant to the hadronic component over all radii and energy bands
considered.

The predicted enclosed IC fluxes are summarized in
Table~\ref{tab:ic_flux_variation}. For the baseline model
($X_{\rm CRe}=10^{-5}$,
$X_{\rm CRp}=10^{-2}$,
$\alpha_p=2.4$), the integrated IC flux in the broad
$100~\mathrm{MeV}$--$100~\mathrm{TeV}$ band increases from
$7.56\times10^{-12}\ \mathrm{cm^{-2}\,s^{-1}}$
within $R_{500}$ to
$4.86\times10^{-11}\ \mathrm{cm^{-2}\,s^{-1}}$
within $R_{\rm trunc}$.
In the
$1$--$300~\mathrm{GeV}$ energy range adopted for our Fermi-LAT
analysis, the enclosed IC flux increases from
$5.12\times10^{-13}\ \mathrm{cm^{-2}\,s^{-1}}$
to
$3.24\times10^{-12}\ \mathrm{cm^{-2}\,s^{-1}}$,
remaining almost two orders of magnitude below both the predicted
hadronic emission and the observed Fermi-LAT flux of
$1.3\times10^{-10}\ \mathrm{cm^{-2}\,s^{-1}}$~\cite{Manna2024}.

The enclosed IC flux depends on both the primary electron energy
fraction and the parent CR proton spectrum. Increasing the primary
electron energy fraction from
$X_{\rm CRe}=10^{-6}$ to
$10^{-4}$ increases the primary IC flux by two orders of magnitude,
consistent with the nearly linear dependence of the IC luminosity on
the primary electron energy normalization. 
In contrast, varying the proton
spectral index changes the secondary IC emission produced through
hadronic interactions. Over the range
$\alpha_p=2.2$--2.6, the
$1$--$300~\mathrm{GeV}$ enclosed flux within
$R_{\rm trunc}$ decreases from
$8.98\times10^{-12}$ to
$5.02\times10^{-13}\ \mathrm{cm^{-2}\,s^{-1}}$,
demonstrating the strong sensitivity of the secondary IC emission to
the underlying CR proton spectrum. Nevertheless, even for the hardest
proton spectrum considered ($\alpha_p=2.2$), the predicted IC flux
remains more than an order of magnitude below the observed Fermi-LAT
flux, confirming that the detected $\gamma$-ray emission from A3667 is
dominated by hadronic processes rather than IC
scattering.

\begin{table*}[ht]
\centering
\caption{Predicted enclosed IC fluxes from A3667 for
different model parameters. Fluxes are computed using spherical
integration within $R_{500}$ and $R_{\rm trunc}$.}
\label{tab:ic_flux_variation}

\begin{tabular}{llcccc}
\hline
Model &
Parameter &
\multicolumn{2}{c}{$100~\mathrm{MeV}$--$100~\mathrm{TeV}$} &
\multicolumn{2}{c}{$1$--$300~\mathrm{GeV}$} \\

&&
$R_{500}$ &
$R_{\rm trunc}$ &
$R_{500}$ &
$R_{\rm trunc}$ \\

\hline

Baseline &
$X_{\rm CRe}=10^{-5},\ \alpha_p=2.4$
&
$7.56\times10^{-12}$
&
$4.86\times10^{-11}$
&
$5.12\times10^{-13}$
&
$3.24\times10^{-12}$
\\

\hline

Primary CRe &
$X_{\rm CRe}=10^{-6}$
&
$8.97\times10^{-14}$
&
$1.50\times10^{-12}$
&
$5.61\times10^{-15}$
&
$9.35\times10^{-14}$
\\

&
$X_{\rm CRe}=10^{-5}$
&
$8.97\times10^{-13}$
&
$1.50\times10^{-11}$
&
$5.61\times10^{-14}$
&
$9.35\times10^{-13}$
\\

&
$X_{\rm CRe}=10^{-4}$
&
$8.97\times10^{-12}$
&
$1.50\times10^{-10}$
&
$5.61\times10^{-13}$
&
$9.35\times10^{-12}$
\\

\hline

Secondary CRe &
$\alpha_p=2.2$
&
$2.07\times10^{-11}$
&
$1.04\times10^{-10}$
&
$1.78\times10^{-12}$
&
$8.98\times10^{-12}$
\\

&
$\alpha_p=2.4$
&
$6.66\times10^{-12}$
&
$3.36\times10^{-11}$
&
$4.56\times10^{-13}$
&
$2.30\times10^{-12}$
\\

&
$\alpha_p=2.6$
&
$1.83\times10^{-12}$
&
$9.23\times10^{-12}$
&
$9.95\times10^{-14}$
&
$5.02\times10^{-13}$
\\

\hline
\end{tabular}
\end{table*}

\subsubsection{IC Surface Brightness Map}
\label{sec:ic_map}
\begin{figure}[ht]
    \centering
    \includegraphics[width=0.8\columnwidth]{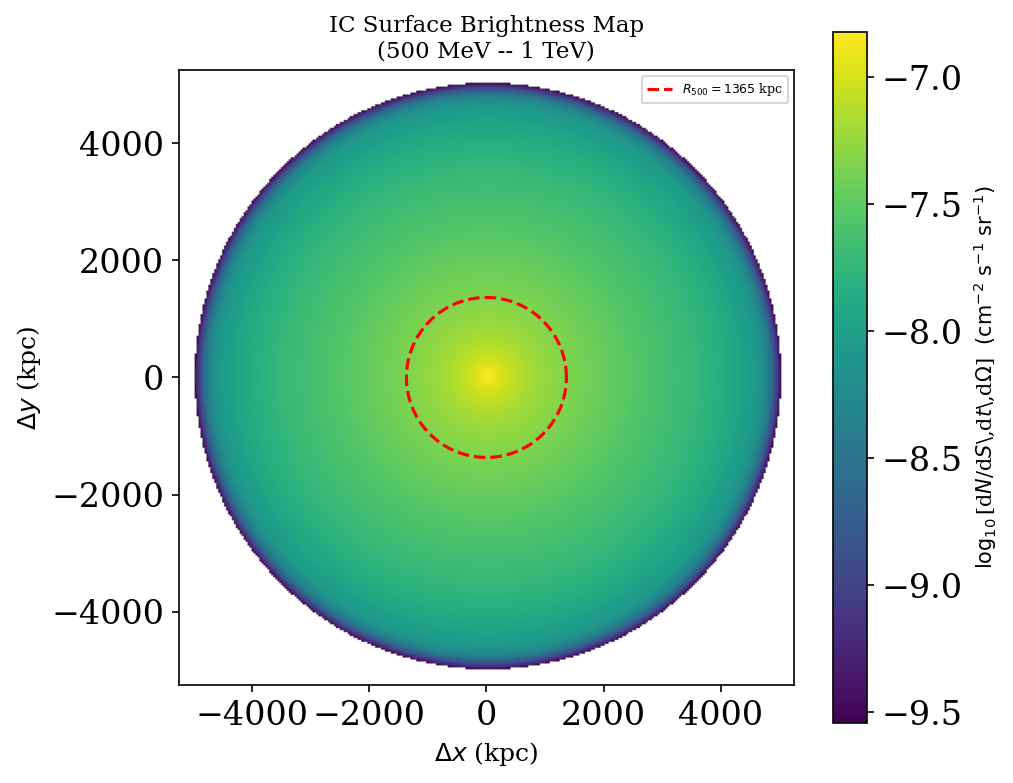}
    \caption{Predicted IC surface brightness map of A3667
    in the $500\,\mathrm{MeV}$--$1\,\mathrm{TeV}$ band,
    shown on a logarithmic colour scale. The red dashed circle
    marks $R_{500} = 1365\,\mathrm{kpc}$. The map is clipped to $\pm R_{\rm trunc}$ in
    both axes. Compared to the hadronic pion-decay map, the
    IC emission is more spatially extended with a shallower
    central concentration, reflecting the linear rather than
    quadratic dependence of the IC emissivity on the thermal
    gas density.}
    \label{fig:ic_map}
\end{figure}

Figure~\ref{fig:ic_map} shows the two-dimensional IC surface
brightness map of A3667 in the $500\,\mathrm{MeV}$--$1\,
\mathrm{TeV}$ band. The map is clipped to $\pm R_{\rm trunc}$
in both axes and displayed on a logarithmic colour scale.

The IC emission is circularly symmetric and spatially
extended, with a shallower central concentration and a
broader morphology than the hadronic pion-decay map of
Section~\ref{sec:gamma_map}. The peak surface brightness at
the cluster center is lower than in the hadronic case.

The red dashed circle marks $R_{500} = 1365\,\mathrm{kpc}$,
corresponding to an angular radius of $\theta_{500} = 0.340^\circ$
at the cluster redshift. The IC emission fills and extends
well beyond this circle, in contrast to the hadronic map
where the brightest emission is tightly concentrated within
$R_{500}$, even though both channels contribute the majority
of their integrated flux from beyond $R_{500}$ (see
Sections~\ref{sec:gamma_flux} and~\ref{sec:ic_flux}).

As with the hadronic map, the circular symmetry of the IC
emission reflects the azimuthally averaged input profiles;
the true IC morphology in A3667 would be influenced by the
actual CRe transport, energy loss history, and the asymmetric
gas distribution associated with this merging system. These
effects are not captured by our spherically symmetric model.

\section{Comparison with Multi-wavelength Observational Constraints}
\label{sec:obs_constraints}

\begin{figure}[ht]
\centering
\includegraphics[width=0.82\columnwidth]{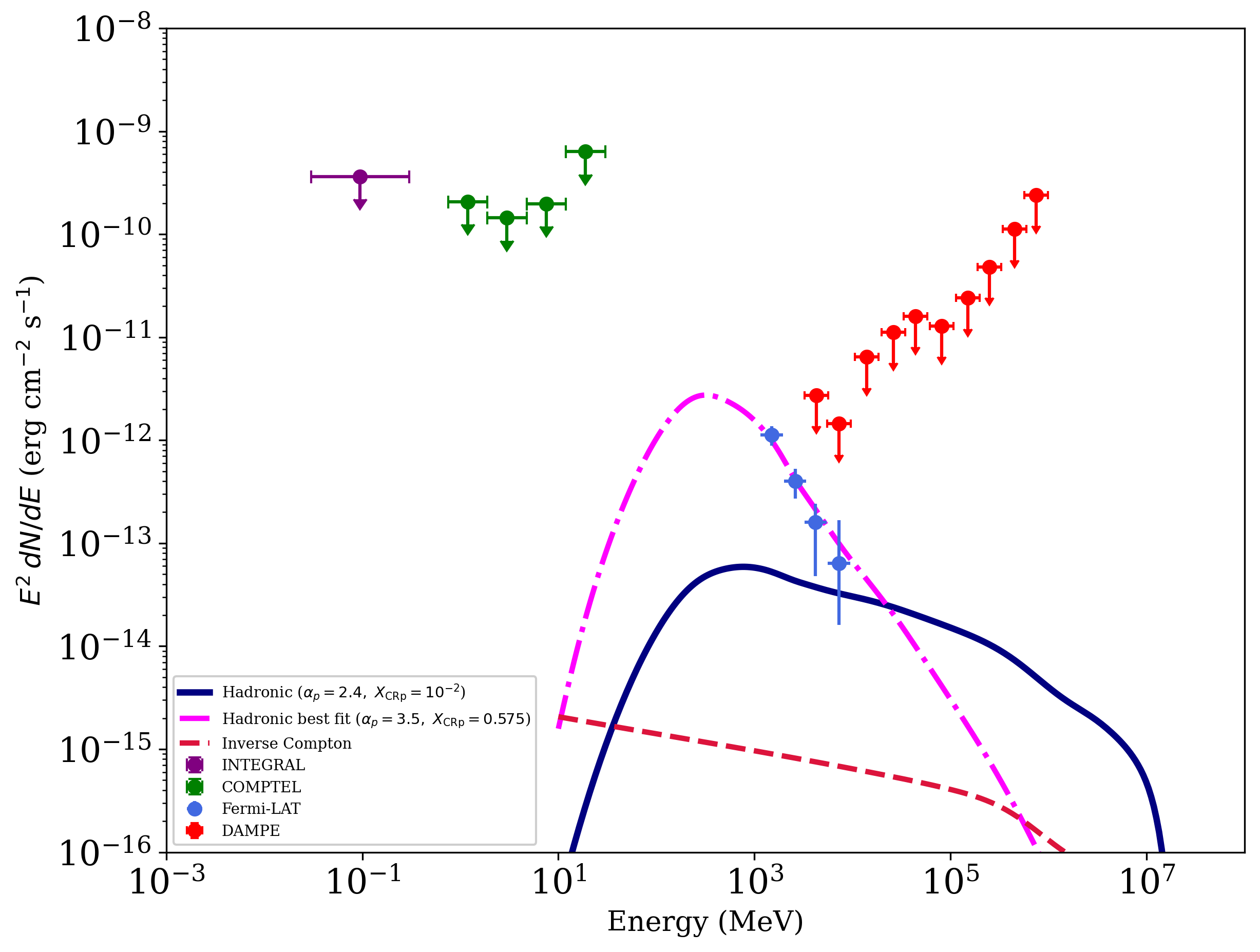}
\caption{Broadband non-thermal spectral energy distribution of
Abell~3667. The solid blue curve shows the baseline hadronic
$\pi^{0}$-decay emission
($X_{\rm CRp}=10^{-2}$,
$X_{\rm CRe}=10^{-5}$,
$\alpha_p=2.4$,
$\eta=1$), and the dashed red curve shows the
IC emission from cosmic-ray electrons under the same baseline
model. The magenta dash-dotted curve shows the hadronic emission for
the best-fit solution at fixed $\alpha_p=3.5$
($X_{\rm CRp}=0.575$; see Section~\ref{sec:gamma_spectrum}), which
provides a substantially improved match to the Fermi-LAT data
relative to the baseline model. Also shown are
the reported Fermi-LAT spectral measurements together with
published upper limits from COMPTEL, INTEGRAL/IBIS--ISGRI, and DAMPE. Both models remain below all
currently available observational upper limits, demonstrating
consistency with existing multi-wavelength constraints.}
\label{fig:sed_constraints}
\end{figure}

Figure~\ref{fig:sed_constraints} compares the predicted broadband
non-thermal spectral energy distribution (SED) of Abell~3667 with the
observational constraints spanning the hard
X-ray to TeV $\gamma$-ray regime. The model includes both the
hadronic $\pi^{0}$-decay component produced by cosmic-ray proton
interactions and the IC emission from cosmic-ray
electrons calculated using the fiducial MINOT parameters
($X_{\rm CRp}=10^{-2}$,
$X_{\rm CRe}=10^{-5}$,
$\alpha_p=2.4$,
$\eta=1$). The predicted hadronic component dominates the non-thermal emission
over the entire energy range considered, whereas the IC contribution
remains more than an order of magnitude lower throughout the
$\gamma$-ray band.

The observational data include the reported Fermi-LAT (1--300 GeV)
spectral measurements of Abell~3667~\cite{Manna2024} with upper limits from
COMPTEL (0.75--30 MeV)~\cite{Manna2024b}, INTEGRAL/IBIS--ISGRI (30--300 keV)~\cite{Manna2025i}, and DAMPE (3--1000 GeV)~\cite{Manna2024dampe}. The baseline MINOT model lies below all currently available upper
limits from these instruments, indicating that it is fully consistent
with existing observational constraints.

\section{Discussion and Conclusions}
\label{sec:conclusions}

In this work,  we have investigated the non-thermal $\gamma$-ray
emission expected from the merging galaxy cluster A3667 using the
\textsc{MINOT} framework \citep{Adam2020}. The calculations combine
the observed thermodynamic properties of the intracluster medium
derived from archival \textit{Chandra} observations
\citep{Cavagnolo2009} with physically motivated models for hadronic
and inverse-Compton (IC) emission from cosmic-ray (CR) particles.
Motivated by the Fermi-LAT detection reported in a recent work~\cite{Manna2024}, we further investigated how the predicted emission
depends on the assumed CR proton and electron populations. Our main
results are summarized below.

\begin{enumerate}

\item \textbf{Hadronic emission dominates the predicted
$\gamma$-ray signal.}
For the fiducial model
($X_{\rm CRp}=10^{-2}$,
$\alpha_p=2.4$,
$\eta=1$),
the predicted hadronic flux in the
$1$--$300~\mathrm{GeV}$ band is
$2.82\times10^{-11}\ \mathrm{cm^{-2}\,s^{-1}}$
within $R_{500}$ and
$1.15\times10^{-10}\ \mathrm{cm^{-2}\,s^{-1}}$
within the truncation radius. The latter is in excellent agreement
with the Fermi-LAT integrated flux of
$1.3\times10^{-10}\ \mathrm{cm^{-2}\,s^{-1}}$
reported by \citet{Manna2024}. Approximately
$76\%$ of the predicted hadronic emission originates outside
$R_{500}$, demonstrating that the cluster outskirts contribute
significantly to the total $\gamma$-ray luminosity.

\item \textbf{The IC component remains subdominant.}
The predicted IC emission from both primary and secondary cosmic-ray
electrons is more than an order of magnitude below the hadronic
component throughout the Fermi-LAT energy range. For the fiducial
model, the integrated IC flux within
$R_{\rm trunc}$ is only
$3.24\times10^{-12}\ \mathrm{cm^{-2}\,s^{-1}}$
between $1$ and $300~\mathrm{GeV}$, confirming that IC scattering
cannot account for the reported Fermi-LAT signal. Although the IC
emission is more spatially extended than the hadronic component, it
remains undetectable with current $\gamma$-ray instrumentation.

\item \textbf{The predicted emission is strongly dependent on the CR
energy density and spectral slope.}
Varying the CR proton energy fraction changes the hadronic flux
almost linearly over two orders of magnitude, whereas varying the
proton spectral index primarily redistributes the emitted power over
energy. Similarly, the normalization of the primary IC component is
controlled by the primary electron energy fraction, while the
secondary IC emission is governed by the parent CR proton spectrum.
These parameter studies demonstrate that the commonly adopted values
$X_{\rm CRp}=10^{-2}$ and
$X_{\rm CRe}=10^{-5}$ should be regarded as fiducial reference
parameters rather than uniquely determined physical quantities.

\item \textbf{Comparison with the Fermi-LAT spectrum constrains the CR
population.}
\rthis{We have shown that the Fermi-LAT detected flux and spectral slope ($\Gamma= 3.61\pm 0.33$) can be well accounted for by hadronic emission assuming a proton spectrum $\propto E^{-\alpha}$ with $\alpha \simeq 3.5-3.8$. At the same time, the energy density of accelerated protons should be $\sim 0.5-1$ times the energy density of thermal plasma, pointing toward the equipartition between the thermal and non-thermal component. Even though the spectral slope seems very steep, they can be accounted for by acceleration at accretion shocks in the cluster with low Mach number $\sim 2$.}

\item \textbf{Consistency with multi-wavelength observational constraints.}
A comparison of the predicted broadband non-thermal spectral energy
distribution with the available COMPTEL, INTEGRAL/IBIS--ISGRI,
DAMPE, and Fermi-LAT observations shows that the baseline
MINOT model remains below all currently available upper limits from
hard X-ray to TeV energies. The hadronic component dominates the
predicted emission over the entire $\gamma$-ray band, while the
inverse-Compton contribution remains subdominant. This comparison,
presented in Section~\ref{sec:obs_constraints}, demonstrates that the
fiducial cosmic-ray model is consistent with the current
multi-wavelength observational constraints.

\end{enumerate}

The present calculations assume spherical symmetry together with
azimuthally averaged gas density, temperature, magnetic-field and CR
distributions. However, A3667 is a dynamically disturbed merging
cluster exhibiting prominent double radio relics, cold fronts,  and a
highly asymmetric intracluster medium
\citep{Vikhlinin2001,Owers2009,JohnstonHollitt2008}. A more realistic
treatment should therefore incorporate spatially resolved CR
acceleration at the merger shocks together with anisotropic transport
and magnetic-field structure.

Future observations with next-generation facilities such as the
Cherenkov Telescope Array Observatory (CTAO) will provide
substantially improved sensitivity over the
$20~\mathrm{GeV}$--$300~\mathrm{TeV}$ energy range and will be well
suited for testing the hadronic origin of the $\gamma$-ray emission
predicted in this work. Joint analyses combining CTAO observations
with deep radio and X-ray data will enable simultaneous constraints on
the CR proton energy density, electron population, magnetic-field
distribution, and particle acceleration mechanisms in the intracluster
medium. Future work will also extend the present analysis by
incorporating self-consistent cosmic-ray transport and reacceleration
processes, allowing the effects of spatial diffusion,
energy-dependent transport, and shock reacceleration on the predicted
non-thermal emission to be quantified. In addition, a comprehensive
multi-wavelength comparison including the observational upper limits
from instruments such as COMPTEL, INTEGRAL, and DAMPE, together with
future $\gamma$-ray observations, will provide more stringent
constraints on the allowed cosmic-ray parameter space and the physical
origin of the high-energy emission from A3667.

\section{Acknowledgements}
The authors sincerely thank the anonymous referee for the careful reading of the manuscript and for the constructive comments and insightful suggestions, which have significantly improved the quality, clarity, and scientific content of this work. SM gratefully acknowledges the Ministry of Education (MoE), Government of India, for its consistent financial support through the research fellowship, which has been instrumental in facilitating the successful completion of this work. 

\clearpage
\bibliography{references}
\newpage

\appendix
\section{Derived Thermodynamic Profiles of A3667}
\label{app:thermodynamic}
This appendix presents the full set of derived thermodynamic
and structural profiles of A3667 obtained from the density and pressure models described in Section~\ref{sec:profiles}. The electron density
($\beta$-model), thermal pressure ($P = n_e \times T$), and temperature (power-law) profiles are shown individually in the main text
(Figure~\ref{fig:combined_profiles});
the six derived quantities discussed below are collected in
Figure~\ref{fig:thermodynamic_6panel}. All profiles are
evaluated on a logarithmic radial grid spanning
$10$--$5000\ \mathrm{kpc}$.

\subsection{Entropy Profile}
\label{app:entropy}
The ICM entropy is shown in panel~(a) of Figure~\ref{fig:thermodynamic_6panel}. The
profile increases monotonically and confirms that the pressure model satisfies the thermodynamic requirement $\mathrm{d}K/\mathrm{d}r > 0$ \citep{Voit2005} across all radii.

\subsection{Enclosed Thermal Energy}
\label{app:thermal_energy}
\begin{figure*}[ht]
\centering
\includegraphics[width=\textwidth]{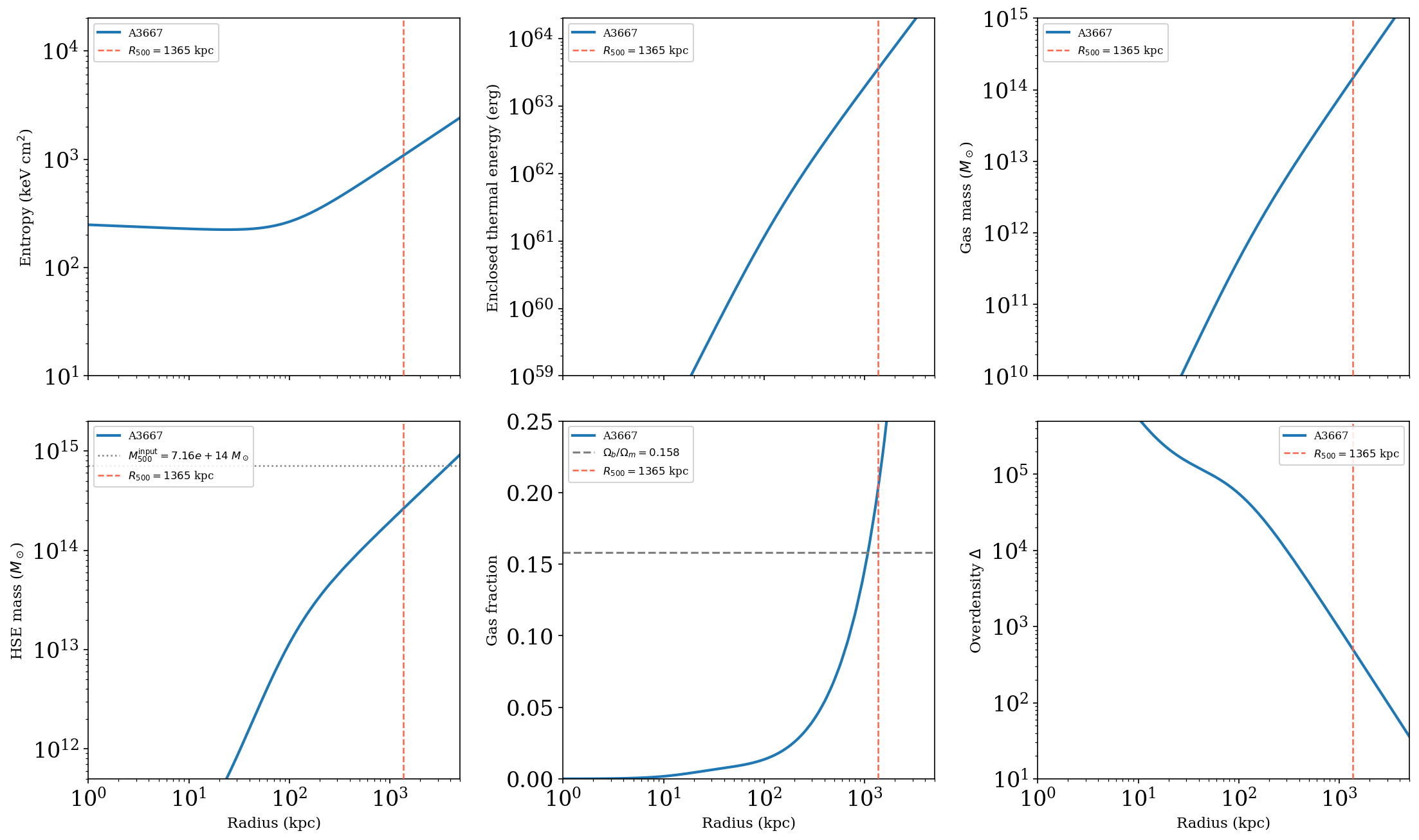}
\caption{Derived thermodynamic profiles of A3667 evaluated
from the $\beta$-model density and power-law
temperature profiles over $10$--$5000\ \mathrm{kpc}$.
\textit{(a)}~ICM entropy.
\textit{(b)}~Enclosed thermal energy $U_{\rm th}(<r)$.
\textit{(c)}~Enclosed gas mass $M_{\rm gas}(<r)$.
\textit{(d)}~Hydrostatic mass $M_{\rm HSE}(<r)$.
\textit{(e)}~Cumulative gas fraction $f_{\rm gas}(<r) =
M_{\rm gas}/M_{\rm HSE}$; the horizontal dashed line marks
the cosmic baryon fraction $f_b = 0.158$ \citep{Planck2015}.
\textit{(f)}~Overdensity contrast $\Delta(r)$; the vertical
dashed line marks $R_{500}^{\rm HSE}$ and the horizontal
dotted line marks $\Delta = 500$.
In all panels the vertical red dashed line marks
$R_{500} = 1365.5\ \mathrm{kpc}$ from the SPT-SZ mass. All
profiles are model-dependent beyond $r \approx 295\
\mathrm{kpc}$ ($0.22\,R_{500}$), the outermost radius of the
ACCEPT data.}
\label{fig:thermodynamic_6panel}
\end{figure*}

The enclosed thermal energy $U_{\rm th}(<r)$ is shown in
panel~(b) of Figure~\ref{fig:thermodynamic_6panel}. The
profile reaches $U_{\rm th}(<R_{500}) \sim 10^{63}\
\mathrm{erg}$, typical for a cluster of this mass at low
redshift \citep{Arnaud2010}.

\subsection{Gas Mass Profile}
\label{app:gas_mass}
The enclosed gas mass $M_{\rm gas}(<r)$ is shown in panel~(c). The gas fraction $f_{\rm gas} = M_{\rm gas}/M_{500}$ approaches
${\sim}0.15$--$0.18$ near $R_{500}$, broadly consistent with
the cosmic baryon fraction $f_b = \Omega_b/\Omega_m = 0.158$
\citep{Planck2015}, as shown in panel~(e).

\subsection{Hydrostatic Mass Profile and HSE Bias}
\label{app:hse_mass}
The hydrostatic equilibrium (HSE) mass profile
$M_{\rm HSE}(<r)$, shown in panel~(d), is derived
analytically from the X-ray temperature and density profiles
under the assumption of spherical symmetry and thermal
pressure support. The ACCEPT thermodynamic data extend only to
$0.216\,R_{500} = 295\ \mathrm{kpc}$. The evaluation of
$M_{\rm HSE}(R_{500})$ therefore relies entirely on the
extrapolated $\beta$-model density and power-law temperature
profiles beyond the data boundary. The result is
model-dependent and should not be interpreted as a direct
observational constraint. 

\subsection{Gas Fraction Profile}
\label{app:fgas}
The cumulative gas fraction $f_{\rm gas}(<r) =
M_{\rm gas}(<r)/M_{\rm HSE}(<r)$ is shown in panel~(e), with
the cosmic baryon fraction $f_b = \Omega_b/\Omega_m = 0.158$
\citep{Planck2015} overlaid as a reference. The gas fraction
rises steeply at small radii and approaches the cosmic baryon
fraction near $R_{500}$. 

\subsection{Overdensity Profile}
\label{app:overdensity}
The mean overdensity contrast $\Delta(r) \equiv
\bar{\rho}(<r)/\rho_c(z)$, where $\bar{\rho}(<r) =
3\,M_{\rm HSE}(<r)/(4\pi r^3)$ is the mean enclosed mass
density, is shown in panel~(f). The radius at which
$\Delta = 500$ defines $R_{500}^{\rm HSE}$ from the profile,
indicated by the dashed vertical line. This value is
consistent with the input $R_{500} = 1365.5\ \mathrm{kpc}$
derived from the SPT-SZ mass
(Section~\ref{sec:cluster_props}), confirming internal
consistency of the thermodynamic modeling.

\end{document}